\documentclass[reprint]{revtex4-1}

\usepackage{longtable}

\usepackage{natbib}

\usepackage{graphicx}
\usepackage{amsmath}
\usepackage{amsfonts}
\usepackage{color}

\begin{document}


\title{Deformation of an Amorphous Polymer during the \\ Fused-Filament-Fabrication Method for Additive Manufacturing} 



\author{Claire McIlroy}
\email[]{cm1509@georgetown.edu}
\affiliation{Department of Physics and Institute for Soft Matter Synthesis and Metrology Georgetown University, Washington DC, USA}

\author{Peter D. Olmsted}
\email[]{pdo7@georgetown.edu}
\affiliation{Department of Physics and Institute for Soft Matter Synthesis and Metrology Georgetown University, Washington DC, USA}


\date{\today}

\begin{abstract}
3D printing is rapidly becoming an effective means of prototyping and creating custom consumer goods. The most common method for printing a polymer melt is fused filament fabrication (FFF), and involves extrusion of a thermoplastic material through a heated nozzle; the material is then built up layer-by-layer to fabricate a three-dimensional object.
Under typical printing conditions the melt experiences high strain rates within the FFF nozzle, which are able to significantly stretch and orient the polymer molecules. In this paper, we model the deformation of an amorphous polymer melt during the extrusion process, where the fluid must make a 90$^\text{o}$ turn. The melt is described by a modified version of the Rolie-Poly model, which allows for flow-induced changes in the entanglement density. The complex polymer configurations in the cross-section of a printed layer are quantified and visualised. The deposition process involving the corner flow geometry dominates the deformation and significantly disentangles the melt.  
\end{abstract}

\pacs{}

\maketitle 




%
%

%

\section*{Nomenclature}

\begin{longtable}[l]{p{50pt} p{200pt}}
\addtocounter{table}{-1}
\textbf{Symbol}	& \textbf{Description} \\
$t$ & Time \\
$T$ & Temperature \\
$R$ & Nozzle outlet radius \\
$R_0$ & Radius of heated nozzle section \\
$L$ & Nozzle length \\
$L_0$ & Length of heated nozzle section \\
$H$ & Layer thickness \\
$Q$ & Mass flow rate \\
${\bf u}$ & Velocity vector \\
$U$ & Magnitude of velocity vector \\
${\bf K} $ & Velocity gradient tensor \\
$\dot{\gamma}$ & Shear rate \\
$\dot{\gamma}_W$ & Shear rate at nozzle wall\\
$\overline{Wi}_N$ & Mass-averaged equilibrium reptation Weissenberg number \\
$\overline{Wi}_N^R$ & Mass-averaged equilibrium Rouse Weissenberg number \\
$Wi_W$ & Equilibrium reptation Weisseberg number at nozzle wall \\
$Wi_W^R$ & Equilibrium Rouse Weisseberg number at nozzle wall \\
$U_N$	 	&  Vertical average print speed \\
$U_L$       &  Horizontal average print speed \\
Re & Reynolds' number \\
$p$ & Pressure \\
$\rho$ & Mass density \\
$G_e$ & Plateau modulus \\
$\mu_s$ & Rouse viscosity \\
$Z_{eq}$ & Entanglement number \\
$M_w$     & Molecular weight \\
$M_e$     & Entanglement molecular weight \\
$T_0$    & Reference temperature \\
$T_N$ & Print temperature \\
$a(T)$   & WLF shift factor \\
$C_1$ & WLF parameter \\
$C_2$ & WLF parameter \\
$\alpha$ & Thermal diffusivity \\
$\tau_e^0$ & Rouse time of one entanglement segment at $T_0$ \\
$\tau_R^0$ & Rouse time of polymer chain at $T_0$\\
$\tau_d^0$ & Reptation time of chain at $T_0$ \\
$\tau_{res}$ & Residence time in nozzle \\
$\tau_{dep}$ & Deposition time  \\
$L_{skin}$ & Thermal skin layer in deposit \\
$\tau_{sw}$ & Die swell time scale \\
$z_M$ & Terminal swell distance \\
$\boldsymbol{\sigma}$ & Total stress \\
$\bf A$ & Polymer deformation tensor \\
$\bf R$ & Polymer end-to-end vector \\
$R_g$ & Polymer radius of gyration \\
$A_{rs}$ & Principle shear deformation \\
$\text{tr}{\bf A} -3$ & Stretch deformation \\
$N$ & Normal stress difference \\
$\lambda_1$ & Principle eigenvalue \\
$\hat{\bf e}_1$ & Principle eigenvector \\
$\eta_\theta$ & Polar angle \\
$\eta_\phi$ & Azimuthal angle \\
$\nu$ & Entanglement fraction \\
$\beta$ & Convective constraint release parameter \\
$\nu_N$ & Entanglement fraction at nozzle wall \\
$\nu_L$ & Entanglement fraction at weld site \\
$\hat{\bf s}$ & Flow direction \\
$\hat{\bf e}_x$, $\hat{\bf e}_y, \hat{\bf e}_z$ & Orthonormal Cartesian coordinate basis \\
$\hat{\bf r}, \hat{\boldsymbol{\theta}}, \hat{\boldsymbol{\phi}}$ & Orthonormal spherical coordinate basis \\
$r^0, \phi^0$ & Initial polar coordinates \\
$r, \phi$ & Transformed polar coordinates \\
$\theta$ & Polar angle between nozzle and layer \\
$\boldsymbol{\mathcal{R}}$ & Mesh points \\
$\bf M$ & Rotation matrix \\
$\boldsymbol{\Lambda}$ & Deformation factor tensor \\
$\boldsymbol{\Omega}$ & Rotation Matrix \\
$\boldsymbol{\mathcal{M}}$ & Transformation matrix \\
$d {\mathcal{A}}$ & Area of deposition cross-section \\
$da$ & Area of mesh element \\
$ds$ & Displacement \\
$\Delta t$ & Advection time 
\end{longtable}

\newpage

\section{Introduction}

Fused filament fabrication (FFF), also know as fused deposition modelling (FDM) \cite{Chua:2003}, is a powerful additive-manufacturing tool. The simple-to-use technology allows the fabrication of complex geometries via build-preparation software, as well as printed parts with locally controlled properties such as density and porosity \citep{Li:2002}. FFF is now considered indispensable for the rapid manufacturing of concept models, functional prototypes and customized end-use parts.

\begin{figure}[t]
\centerline{\includegraphics[width=9cm]{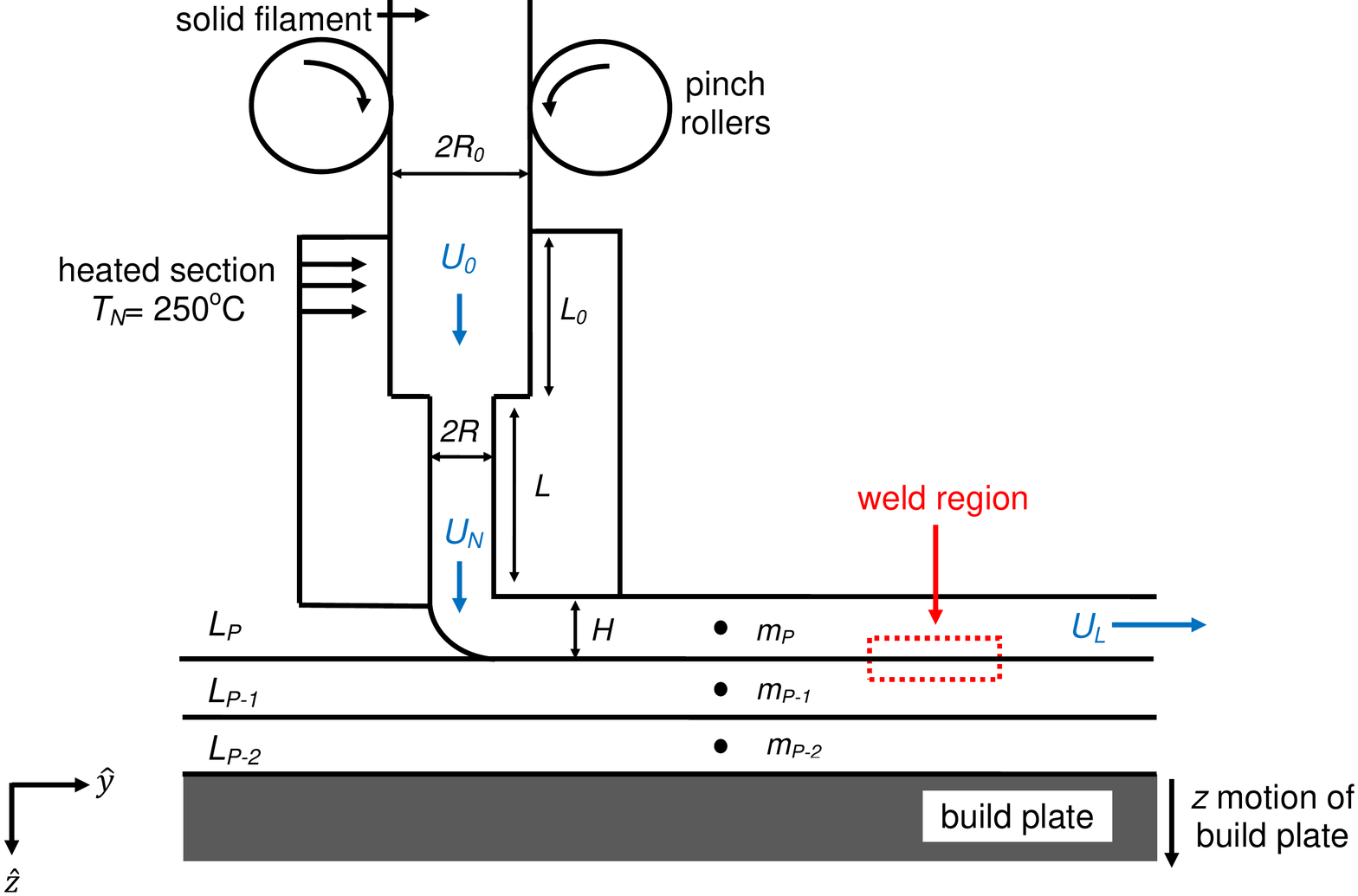}}
\caption{Simple schematic of typical FFF process, as described in text. In frame of moving nozzle, the melt exits the nozzle at speed $U_N$ and the build plate moves at speed $U_L$ in the $\hat{y}$-direction. The current printed layer is denoted $\text{L}_\text{p}$ and the middle of the layer is denoted $\text{m}_\text{p}$. Welds occur at the interface between  layers. The layer thickness $H$ is typically less than the nozzle diameter $2R$. See Appendix \ref{sec:appendix2} for typical model parameters.}
\label{fig:process}
\end{figure}

In FFF, a solid thermoplastic filament is fed into a machine via a pinch-roller mechanism, as shown by the simplified schematic in Fig. \ref{fig:process}. FFF systems contain multiple contractions between the pinch roller and the nozzle exit; for simplicity Fig. \ref{fig:process} shows a single contraction between the heated and final sections of the nozzle. The most common printing material investigated is acrylonitrile butadiene styrene (ABS), an amorphous polymer melt containing  rubber (butadiene) nano-particles \cite{Ziemian:2012}. FFF machines can also print parts from amorphous polycarbonate \citep{Hill:2014}, semi-crystalline poly-lactic acid \citep{Drummer:2012} and other thermoplastic materials \citep{Turner:2014}.

The feedstock is melted and extruded through a nozzle, with the solid portion of the filament acting as a piston to push the melt through. A three-dimensional object is constructed by printing the extrudate layer-by-layer onto a build plate. As the material is deposited the nozzle moves in the $xy$-plane to create a prescribed pattern, and the platform moves in the $z$-direction for additional layers to be built. The thickness of the single layer is determined by the height of the nozzle with respect to the previously printed layer (Fig. \ref{fig:process}), whereas the width of the layer is determined by a combination of the flow rate, surface tension and viscoelasticity. The speed of the material flowing through the nozzle is controlled to prevent drawing and buckling, so that the width of the layer is approximately equal to the nozzle diameter.

Upon deposition, the melt bonds, cools and solidifies with adjoining material so that the structure of the final object consists of a number of partially-bonded filaments. Much of the literature to date has focused on how FFF parameters, such as build style, raster width and raster angle, affect the material properties \citep{Ahn:2002, Ahn:2003, Arivazhagan:2012, Lee:2007}. Analytical models, based on classical lamination theory combined with the Tsai-Wu failure criterion, are used to predict the tensile strength of the bond \citep{Ahn:2003}, and adhesives can be used to alter the bonding behaviour \citep{Esplain:2010}. 

In the absence of a post-deposition cross-linking process, bonding between layers is thermally driven and significantly affected by print temperature \citep{Sun:2008}; higher temperatures can enable better adhesion between printed layers and therefore stronger mechanical strength of the final printed part, while temperatures that are too high lead to polymer degradation and weakening of the product \citep{Gibson:2010}. Recently, carefully-calibrated infra-red imaging has been developed to extract the temperature profile at the weld \citep{Seppala:2016} and finite-element analysis has been used to examine temperature gradients at the nozzle exit \citep{Yardimici:1997}. Laser-assisted heating is proposed to improve the thermal-bonding process and consequently the strength of the printed part \citep{Du:2016}.

A number of studies have investigated the thermal-welding of polymer molecules (e.g. \cite{Wool:1981,Robbins:2013}),
however, these studies focus on the diffusive behaviour of melts in an equilibrium state. 
During FFF the polymer experiences large shear rates in the nozzle and rapid temperature changes that are expected to significantly deform the polymer microstructure. A non-equilibrium microstructure will affect polymer diffusion \cite{Hunt:2009, Uneyama:2011, Ilg:2011} and consequently welding. 
For example, it is suggested that polymer alignment in the flow may lead to de-bonding of the layers and create defects in the final printed object \citep{Sood:2012}.

In this paper, we employ a continuum molecularly-aware polymer model \citep{Likhtman:2003}, which we modify to incorporate flow-induced changes in the entanglement density, to describe the behaviour of a typical amorphous polymeric printing material during the FFF extrusion process.  Assuming steady-state and a uniform temperature profile, the printing flow and polymer configuration tensor within a cylindrical nozzle and during the subsequent deposition are calculated using a simple mapping to represent the $90^\text{o}$ turn and deformation into an elliptical-shaped layer. We quantify and visualise the polymer microstructure across a printed layer post-extrusion and investigate the effect of print speed on disentanglement. The relaxation of this deformation and the effect on weld properties will be considered elsewhere.

\section{Model for FFF}
\label{sec:FFFmodel}

\subsection{Modified Rolie-Poly Model with Flow-Induced Disentanglement}

A linear polymer melt is well described by the Doi-Edwards tube model \citep{DoiEdwards:1986}. In this paper, we implement a variation of this standard theory known as the the Rolie-Poly model \cite{Likhtman:2003} and include a new feature that allows for flow induces changes in the entanglement density \cite{Ianniruberto:2015}. Although the Rolie-Poly model does not allow for second normal stresses in axisymmetric flows, it provides a simple one-mode constitutive equation for the stress tensor to describe entangled polymers.

At equilibrium, the entanglement number of a melt of molecular weight $M_w$ is defined by
\begin{equation}
 Z_{eq} = \frac{M_w}{M_e},
\end{equation}
where $M_e$ is the molecular weight between entanglements. The Rouse and reptation times of a polymer at a given reference temperature $T_0$ are given by \citep{Likhtman:2002}
\begin{subequations}
\begin{align}
 \tau_R^0 &= \tau_e^0 Z_{eq}^2, \\
 \tau_d^0 &= 3 \tau_e^0 Z_{eq}^3 \left(1 - \frac{3.38}{\sqrt{Z_{eq}}} + \frac{4.17}{Z_{eq}} - \frac{1.55}{\sqrt{Z_{eq}}^3} \right),
 \end{align}
\end{subequations}
respectively, where $\tau_e^0$ is the Rouse time of one entanglement segment. 

Due to the non-isothermal conditions of FFF, the temperature-dependent rheology must be considered. We account for this by scaling the relaxation times by a shift factor $a(T)$, typically measured by rheology, which has the well-known WLF form \cite{Williams:1955}
\begin{equation}
 a(T) = \exp \left(\frac{-C_1(T-T_0)}{T+C_2-T_0} \right),
 \label{eq:shift}
\end{equation}
for temperature $T$ and constants $C_1$ and $C_2$. At equilibrium, the Rouse and reptation times of a melt are given by
\begin{equation}
\tau_R^{eq}(T) = \tau_R^0 a(T),
\label{eq:tauReq}
\end{equation}
and
\begin{equation}
 \tau_d^{eq}(T) = \tau_d^0 a(T).
 \label{eq:taudeq}
\end{equation}

\begin{figure}[b!]
\centerline{\includegraphics[width=7cm]{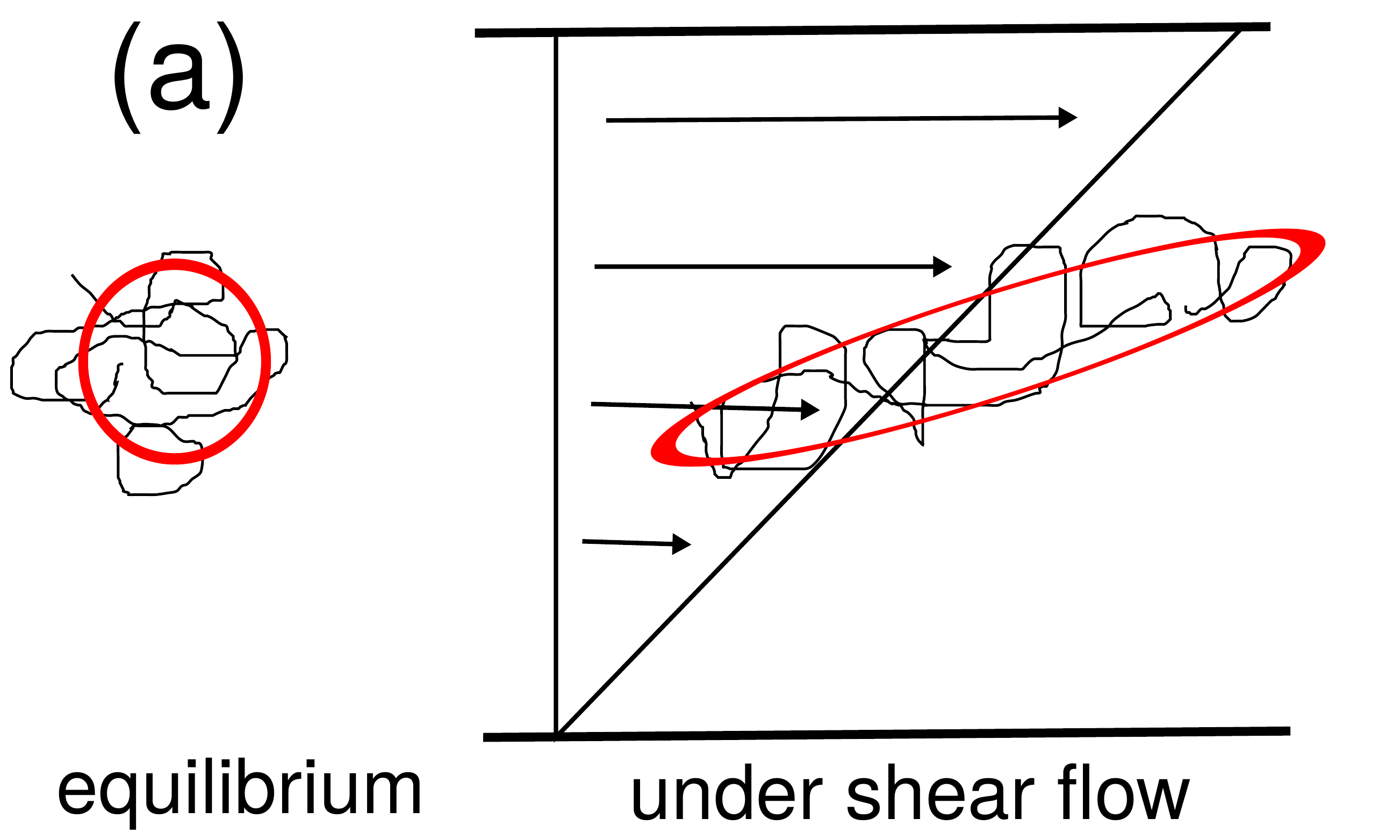}}
 \centerline{\includegraphics[width=7cm]{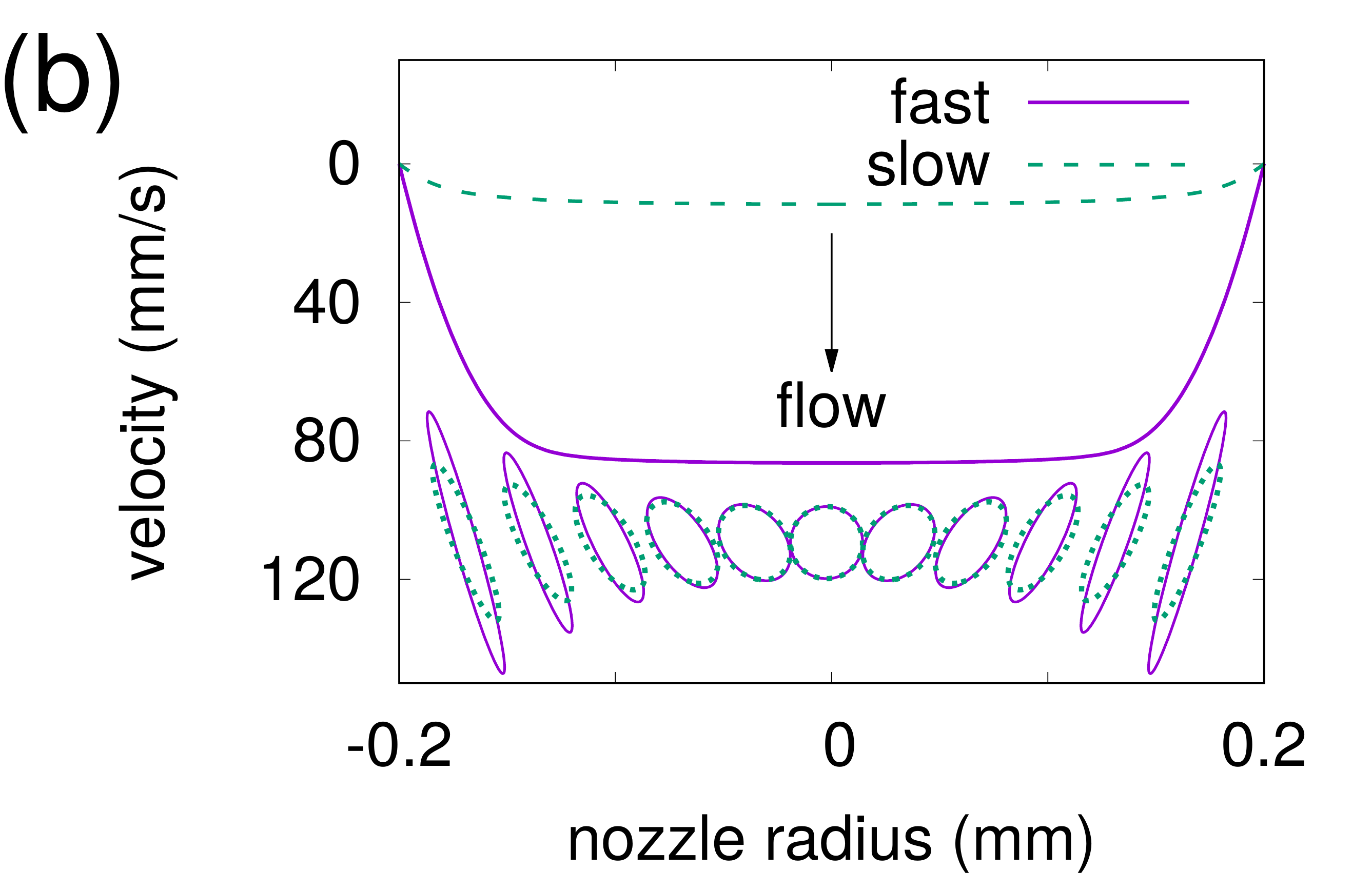}}
\caption{(a) Visualisation of polymer as a deformed sphere under shear flow. (b) Velocity profiles $w(r)$ and ellipsoidal representation of polymer deformation ($Z_{eq}=37, \beta=0.3)$ across the nozzle radius for fast and slow print cases corresponding to $\overline{Wi}_N = 13$ and 2 (Table \ref{tab:Weissenberg}).}
\label{fig:velocity}
\end{figure}

Momentum balance is given by
\begin{equation}
 \rho \frac{D {\bf u}}{Dt} = \nabla \cdot \boldsymbol{\sigma},
\end{equation}
for mass density $\rho$, fluid velocity ${\bf u}$ and the material derivative $\frac{D}{Dt} = \frac{\partial}{\partial t} + ({\bf u}\cdot\nabla)$. In steady state we solve
\begin{equation}
 \nabla \cdot \boldsymbol{\sigma} =0,
 \label{eq:NSsteady}
\end{equation}
for stress tensor $\boldsymbol{\sigma}$. The total stress in the polymer melt comprises solvent and polymer contributions
\begin{equation}
\boldsymbol{\sigma} = p {\bf I} + G_e ({\bf A} - {\bf I}) + 2 \mu_s ( {\bf K} + {\bf K}^T),
\label{eq:stress}
\end{equation}
where $p$ is the isotropic pressure and the velocity gradient tensor is denoted $K_{\alpha \beta} =\nabla_\beta u_{\alpha}$. The polymer contribution to the stress is given by the plateau modulus $G_e$ multiplied by the polymer deformation tensor
\begin{equation}
 {\bf A} = \frac{ \langle {\bf RR} \rangle }{3R_g^2},
\end{equation}
for end-to-end vector ${\bf R}$ and radius of gyration $R_g$. Fig. \ref{fig:velocity}a shows the polymer microstructure, defined by tensor ${\bf A}$, as an ellipsoid; a sphere represents an undeformed polymer at equilibrium $({\bf A}={\bf I})$ with radius $R_g$, whereas an ellipse signifies stretch and orientation. For times shorter than $\tau_e$, Rouse modes corresponding to lengths shorter than $M_e$ contribute to a background viscosity defined as \citep{Graham:thesis}
\begin{equation}
 \mu_s = \frac{\pi^2}{12} \frac{G_e}{Z_{eq}} \tau_R^{eq}.
\end{equation}
We assume that the polymer deformation tensor ${\bf A}$ satisfies the Rolie-Poly equation \citep{Likhtman:2003}
\begin{equation}
\begin{split}
 \frac{D {\bf A}}{Dt} &= {\bf K} \cdot {\bf A} + {\bf A} \cdot {\bf K}^T - \frac{1}{\tau_d(T,\dot{\gamma})} ({\bf A}-{\bf I}) \\ &- \frac{2}{\tau_R(T)} \left( 1 - \sqrt{\frac{3}{\text{tr}{\bf A}}}\right) \left( {\bf A} + \beta \sqrt{\frac{\text{tr}{\bf A}}{3}} ({\bf A}-{\bf I}) \right),
\label{eq:Rolie-Poly}
\end{split}
\end{equation}
where $\text{tr}{\bf A}$ denotes the trace of tensor ${\bf A}$. Convective constraint release (CCR) is incorporated via the parameter $\beta$ \cite{Ianniruberto:1996}. The reptation and Rouse times are denoted $\tau_d(T,\dot{\gamma})$ and $\tau_R(T)$, respectively.

When a melt is subjected to flow, entanglements may be lost via convection and new entanglements made by reptation. We incorporate flow-induced changes in the entanglement fraction $\nu = Z/Z_{eq}$ to the Rolie-Poly model via the recent kinetic equation of Ianniruberto \& Marrucci \citep{Ianniruberto:2014, Ianniruberto:2014erratum, Ianniruberto:2015}:
\begin{equation}
 \frac{D\nu}{Dt} = - \beta \left(  {\bf K}: {\bf A} - \frac{1}{\text{tr}{\bf A}} \frac{d \text{tr}{\bf A}}{dt}  \right) \nu + \frac{1-\nu}{\tau_d^{eq}(T)},
\label{eq:nu}
\end{equation}
where entanglement loss can be modified by varying the CCR parameter $\beta$. The reptation time is given by a thermal contribution plus a convective one determined by the rate of entanglement loss \cite{Ianniruberto:2015}
\begin{equation}
\frac{1}{\tau_d(T,\dot{\gamma})} = \frac{1}{\tau_d^{eq}(T)} + \beta \left ({\bf K}:{\bf A} - \frac{1}{\text{tr}{\bf A}} \frac{d \text{tr}{\bf A}}{dt} \right),
\label{eq:taud}
\end{equation}
where the temperature-dependence of the equilibrium reptation time is given by Eq. \ref{eq:taudeq}. The reptation time therefore implicitly depends on the shear rate $\dot{\gamma}$. The Rouse time does not depend on the local shear rate and is given by Eq. \ref{eq:tauReq} under flow conditions. The steady-state constitutive curve defined by Eqs. \ref{eq:stress}, \ref{eq:Rolie-Poly} and \ref{eq:nu} demonstrates the shear-thinning behaviour typical of FFF-printed materials and is discussed in Appendix \ref{sec:appendix1}; increasing the CCR parameter $\beta$ acts to suppress excess shear-thinning behaviour \cite{Ianniruberto:1996}. 

In steady state shear, Eq. \ref{eq:nu} reduces to 
\begin{equation}
\nu = \frac{1}{1+\beta A_{rs} \dot{\gamma} \tau_d^{eq}},
\label{eq:nusteady}
\end{equation}
where $A_{rs}$ is the shear component of $\bf A$. Larger shear rates impose a greater deformation on the polymer microstructure and the resulting alignment reduces the entanglement fraction. Ianniruberto compared this flow-induced disentanglement theory \cite{Ianniruberto:2015} to molecular dynamics simulations of simple steady shear flow conducted by Baig {\em et al.} \cite{Baig:2010} for a wide range of shear rates $\dot{\gamma}$, finding that $\beta=0.15$ gives the best fit to the entanglement loss data for $Z_{eq}=14$. The theory is also compared to the step strain response experiments reported by Takahashi {\em et al.} \cite{Takahashi:1993} for a polystyrene melt with $Z_{eq}=12$; in this case good agreement is found for $\beta =0.25$. Inhomogeneous disentanglement has also been found in dissipative particle dynamics simulations by Khomami {\em et al.} \cite{Khomami:2015, Khomami:2015b} of polymer melts with $Z_{eq}=13,17$.

In the following we show results for $\beta=0.3$. For our model parameters, the constitutive curve is monotonic and we avoid shear-banding effects. The effect of increasing $\beta$ to unity in our FFF model (as in reference \citep{Likhtman:2003}), which gives the most extreme case of disentanglement, is shown in Section \ref{sec:CCR}.  We discuss the effect of choosing smaller $\beta$ in Appendix \ref{sec:appendix1}.

\begin{table}[t]
\caption{Weissenberg numbers (Eqs. \ref{eq:WiN}-\ref{eq:WiRN}) calculated during extrusion for $Z_{eq}=37$ and $\beta =0.3$ (similar to polycarbonate printing material), print temperature $T_N=250^\text{o}$C and two typical speeds $U_N=75$ and 10 mm/s. See Appendix \ref{sec:appendix2} for further details.}
\centerline{\begin{tabular}{l|c|c|l|c|c}
\hline
\hline 
{\bf Reptation $Wi$} & Fast & Slow & {\bf Rouse $Wi^R$} & Fast & Slow  \\
\hline
\hline
$\overline{Wi}_N$ (average) & 13 & 2 & $\overline{Wi}^R_N$ (average) & 0.07 & 0.009  \\
$Wi_W$ (wall)    & 91 & 24 & $Wi^R_W$ (wall) & 1.5 & 0.4 \\
\hline
\end{tabular}}
\label{tab:Weissenberg}
\end{table}

\subsection{FFF Parameters and the Printing Process}

In the following, Eqs. \ref{eq:NSsteady}, \ref{eq:stress}, \ref{eq:Rolie-Poly} and \ref{eq:nu} are solved to determine the melt behaviour during steady-state extrusion. Extrusion is treated in two stages in a frame fixed with the nozzle, where the build plate moves in the $\hat{y}$ direction. 

First, the melt flows through a fixed, vertically-orientated nozzle at mass-averaged speed $U_N$. The nozzle is circular in shape so that the flow is axisymmetric. The melt is then deposited onto the build surface, which moves horizontally at mass-averaged speed $U_L$. During this deposition, the fluid must speed up and deform to make a 90$^\text{o}$ turn. The layer thickness $H$ is typically less that the nozzle diameter, so that the shape of the layer is roughly elliptical \citep{Sun:2008}. The print temperature $T_N$ is assumed to be uniform across the nozzle radius and throughout the deposition, and we assume that the flow is steady (see Appendix \ref{sec:appendix2} for details).

Assuming mass conservation, the speeds are related by
 \begin{equation}
\pi R^2 U_N =   \frac{\pi RH}{2} U_L,
\label{eq:massconservation}
 \end{equation}
where $R$ is the nozzle outlet radius. As well as local acceleration due to the corner geometry, the flow must also speed up to conserve mass whilst transforming from circular to elliptical geometry. If $U_N <  HU_L/2R$, then too little material is deposited and drawing occurs; similarly, if $U_N > HU_L/R$ then buckling of the printed material occurs.

The equilibrium mass-averaged reptation Weissenberg number in the nozzle is defined by
\begin{equation}
 \overline{Wi}_N = \frac{U_N}{R} \tau_d^{eq}(T_N),
 \label{eq:WiN}
\end{equation}
for equilibrium reptation time $\tau_d^{eq}$ given by Eq. \ref{eq:taudeq}. Values for a typical print temperature and two typical print speeds (fast and slow) are given in Table \ref{tab:Weissenberg}. Since $\overline{Wi}_N \gg1$, we expect a significant orientation of the polymer in the nozzle. The local Weissenberg number at the nozzle wall
\begin{equation}
 Wi_W = \dot{\gamma}_W \tau_d^{eq}(T_N),
 \label{eq:WiW}
\end{equation}
where $\dot{\gamma}_W$ is the wall shear rate, increases by an order of magnitude. This is due to the combination of two effects: first, the linear increase in stress from the centre towards the nozzle wall; and second, the shear-thinning nature of the polymer melt.

Similarly, the equilibrium mass-averaged Rouse Weissenberg number in the nozzle is defined as
\begin{equation}
\overline{Wi}^R_N = \frac{U_N}{R} \tau_R^{eq}(T_N),
\label{eq:WiRN}
\end{equation}
where the equilibrium Rouse time is given by Eq. \ref{eq:tauReq}. For the fast printing case, the local Weissenberg number is $Wi^R_W \sim 1$ near the nozzle wall, which implies stretch of the polymer tube during extrusion.

In the following, we show results for $Z_{eq}=37$, $\beta=0.3$ $T_N=250^o$C and $\overline{Wi}_N=13$ and 2, which are typical values used for FFF of an an amorphous polymeric printing material \cite{Turner:2014}. For comparison, we have chosen model parameters for Bisphenol A Polycarbonate. The full set of model parameters and the assumptions of this model are discussed in Appendix \ref{sec:appendix2}; the model parameters for polycarbonate are given in Table \ref{tab:polycarbonate}, and typical print speeds and nozzle dimensions (corresponding to the simplified schematic in Fig. \ref{fig:process}) are given in Tables \ref{tab:speeds} and \ref{tab:geometry}, respectively.

\section{Steady-State Nozzle Flow}
\label{sec:nozzle}

\subsection{Calculation of Nozzle Flow}

The fluid flows along a direction $\hat{\bf s}$ with arc length coordinate $s$. The flow direction $\hat{\bf s}$ changes when the material exits the nozzle according to
\begin{equation}
\hat{\bf s} \equiv \begin{cases}  \hat{\bf e}_z,& \text{ in the nozzle, } \\
                              \hat{\bf e}_y,& \text{ in deposited layer. }
                \end{cases}
\end{equation}
First, we consider steady-state flow through the circular nozzle in the vertical $\bf \hat{s}$-direction. The reasonable steady-state assumption is discussed in Appendix \ref{sec:appendix2}. In cylindrical polar coordinates $(r,\phi,s )$, the velocity profile is denoted
\begin{equation}
 {\bf u}= w(r) \hat{\bf s}, 
\end{equation}
so that the velocity-gradient tensor is
\begin{equation}
 {\bf K} =  \left( \begin{array}{ccc}
0 & 0 & \frac{\partial w}{\partial r} \\
0 & 0 & 0 \\
0 &  0 & 0 \end{array} \right).
\end{equation}
By Eq. \ref{eq:stress}, the total shear stress is given by
\begin{equation}
 \sigma_{rs} = G_e A_{rs} + \mu_s \frac{\partial w}{\partial r}
 \label{eq:nozstress}
\end{equation}
and satisfies the steady-state momentum balance
\begin{equation}
 \frac{\partial p}{\partial s} = \frac{1}{r} \frac{\partial}{\partial r} (r \sigma_{rs}),
 \label{eq:nozNS}
\end{equation}
for a pressure gradient $\partial p / \partial s$ chosen to induce the mean extrusion velocity 
\begin{equation}
 U_N = \int \frac{w(r)}{\pi R^2} d^2 r.
 \label{eq:meanspeed}
\end{equation}
Finally, the polymer deformation is described by the steady-state Rolie-Poly equation
\begin{equation}
\begin{split}
 {\bf K} &\cdot {\bf A} + {\bf A} \cdot {\bf K}^T - \frac{1}{\tau_d(T,\dot{\gamma})} ({\bf A}-{\bf I}) \\&- \frac{2}{\tau_R(T)} \left( 1 - \sqrt{\frac{3}{\text{tr}{\bf A}}}\right) \left( {\bf A} + \beta \sqrt{ \frac{\text{tr}{\bf A}}{3}} ({\bf A}-{\bf I}) \right) = 0,
 \label{eq:nozRP}
 \end{split}
\end{equation}
where $\text{tr}{\bf A} = A_{ss}+A_{\phi\phi}+A_{rr}$. The reptation time is given by
\begin{equation}
\frac{1}{\tau_d(T,\dot{\gamma})} = \frac{1}{\tau_d^{eq}(T)} + \beta ({\bf K}:{\bf A}) ,
\label{eq:noztaud}
\end{equation}
from Eq. \ref{eq:taud} and the entanglement fraction is given by Eq. \ref{eq:nusteady} for $\dot{\gamma}= \partial w / \partial r$.

\begin{figure}[t]
\centerline{\includegraphics[width=8.5cm]{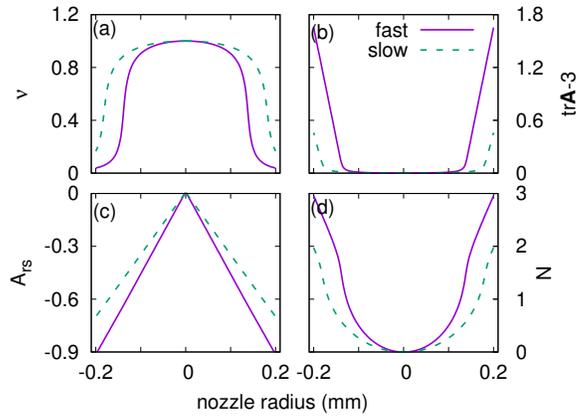}}
\caption{(a-d) Polymer deformation properties in the nozzle $(Z_{eq}=37, \beta=0.3)$: (a) Entanglement fraction profile $\nu(r)$, (b) tube stretch profile $\text{tr}{\bf A}(r)-3$, (c) shear deformation profile $A_{rs}(r)$ and (d) normal stress difference profile $N(r)$. Fast ($\overline{Wi}_N=13$) and slow $(\overline{Wi}_N=2)$ print cases are shown.}
\label{fig:nozdeform}
\end{figure}
\begin{figure}[t!]
\centerline{\includegraphics[width=8.5cm]{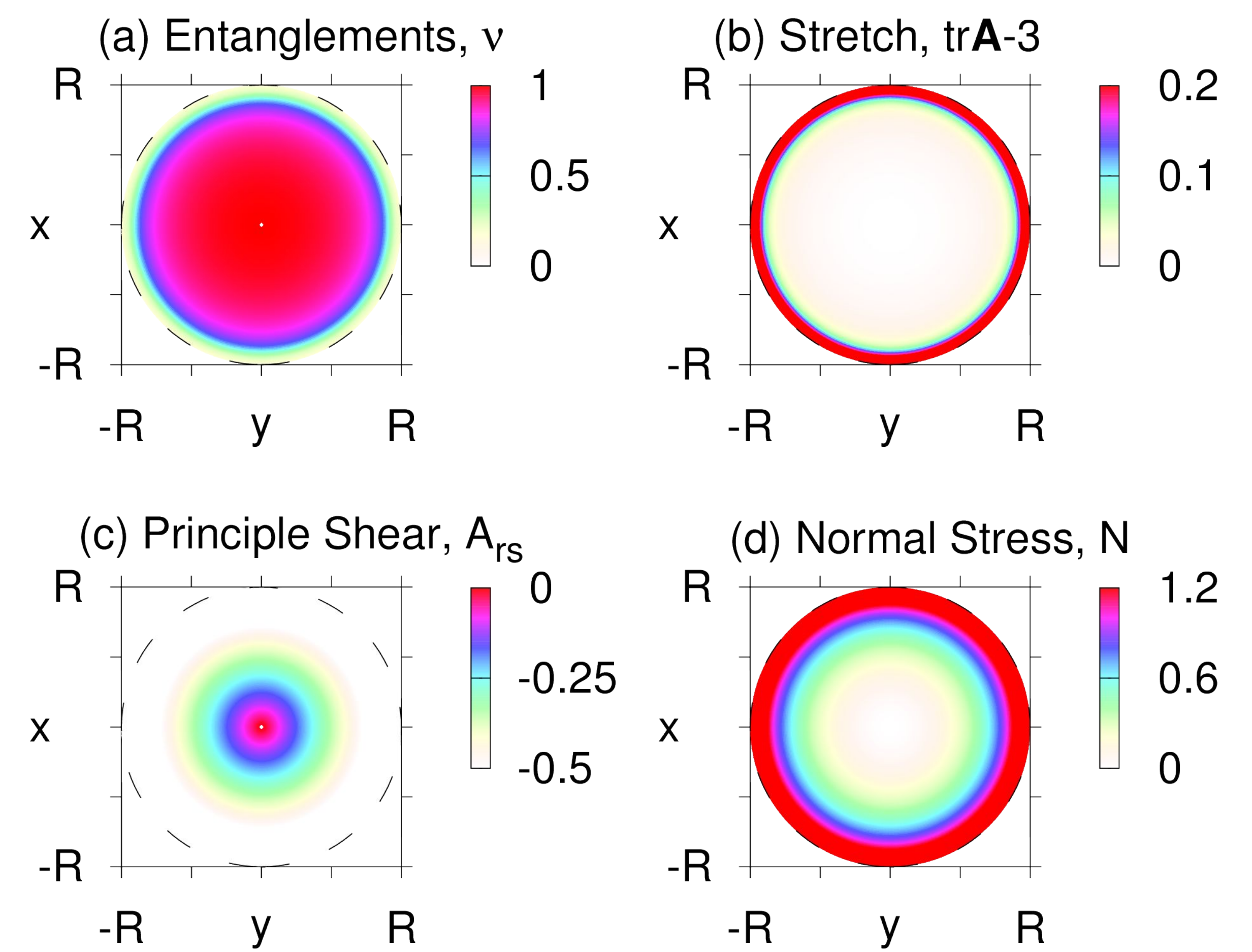}}
\caption{(a-d) Polymer deformation properties in the nozzle for slow-print case $(Z_{eq}=37, \beta=0.3 \text{ and } \overline{Wi}_N=2)$: (a) Entanglement fraction profile $\nu(r,\phi)$, (b) tube stretch $\text{tr}{\bf A}(r,\phi)-3$, (c) principle shear deformation $A_{rs}(r,\phi)$ and (d) normal stress difference $N(r,\phi)$ shown in the $xy$-plane. The fast case $(\overline{Wi}_N=13)$ induces similar deformation profiles.}
\label{fig:nozdeform2}
\end{figure}

\subsection{Polymer Deformation in the Nozzle}

Fig. \ref{fig:velocity} shows the steady-state velocity profiles calculated from Eqs. \ref{eq:nusteady}, \ref{eq:nozstress}-\ref{eq:noztaud} for $Z_{eq}=37$, $\beta=0.3$ and two typical print speeds corresponding to $\overline{Wi}_N=13$ and 2. The profiles have a plug-like shape due to shear-thinning behaviour and are axisymmetric. The ellipses show how the polymer chains becomes more stretched and oriented near the nozzle walls due to the increasing shear rate. 

The polymer deformation for the two typical print speeds is quantified in Figs. \ref{fig:nozdeform}a-d, showing the entanglement fraction $\nu$, the tube stretch $\text{tr}{\bf A}-3$, the shear orientation $A_{rs}$ and the normal stress difference $N=A_{ss}-0.5(A_{rr}+A_{\phi\phi})$. Note that for the Rolie-Poly model, $A_{rr} = A_{\theta\theta}$ in axisymmetric flow, so that $N$ is the first normal stress difference in the nozzle. As expected, the larger Weissenberg number imposes a greater deformation on the polymer, with the chains becoming more stretched and aligned with the flow direction for the fast-print case. Due to this alignment, the entanglement fraction decreases dramatically near the wall (Fig. \ref{fig:nozdeform}a). For $\overline{Wi}_N=2$, $\nu$ is reduced to 20\% of the equilibrium value at the nozzle wall, whereas for $\overline{Wi}_N=13$ the melt is nearly fully disentangled at the wall $(\nu = 5\%)$. These profiles provide an initial condition to calculate ${\bf A}$ during the deposition process.

\begin{figure}[t]
 \begin{minipage}{4cm}
   \centerline{\includegraphics[width=4.5cm]{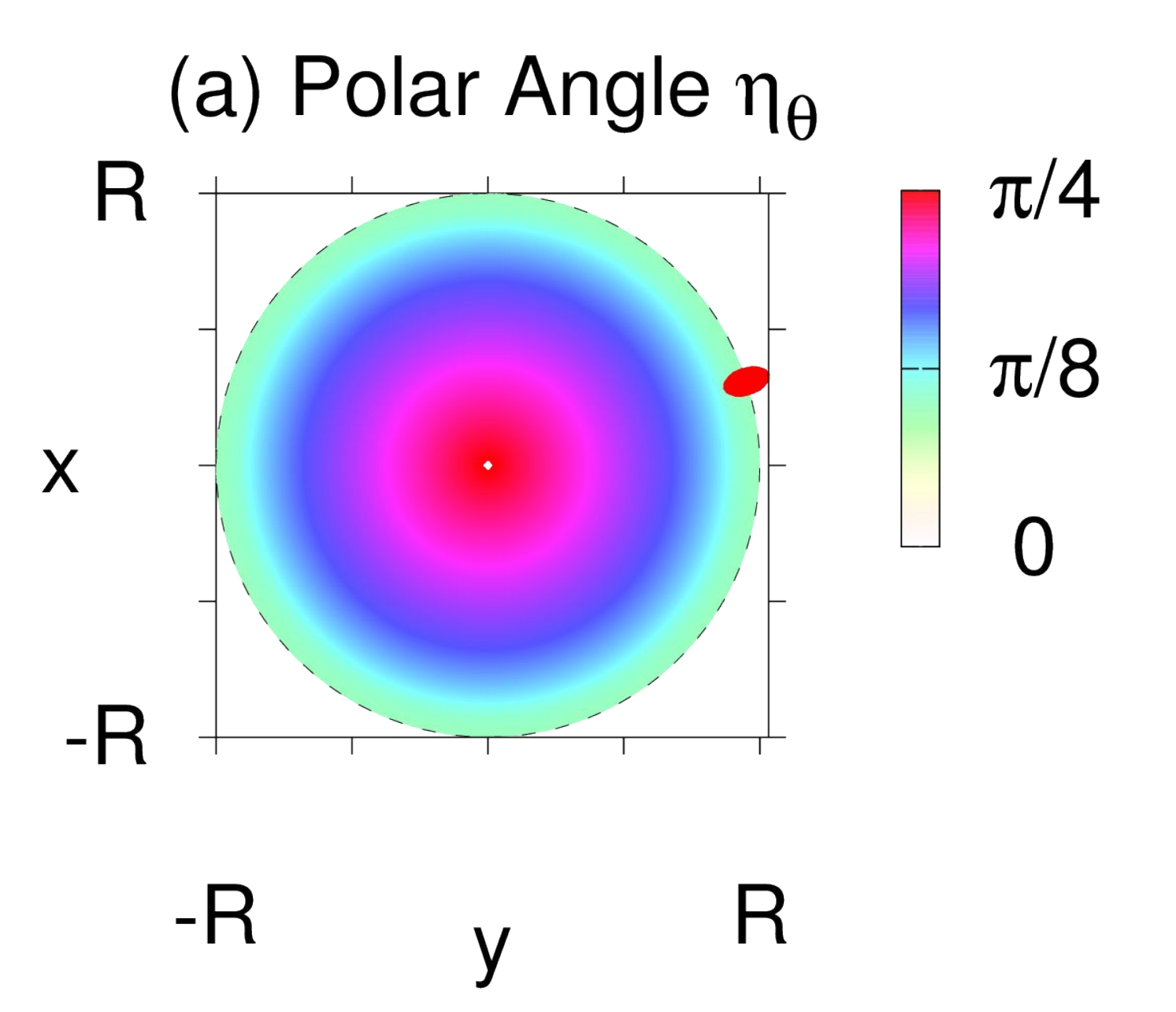}}
 \end{minipage}
 \begin{minipage}{4cm}
   \centerline{\includegraphics[width=4.5cm]{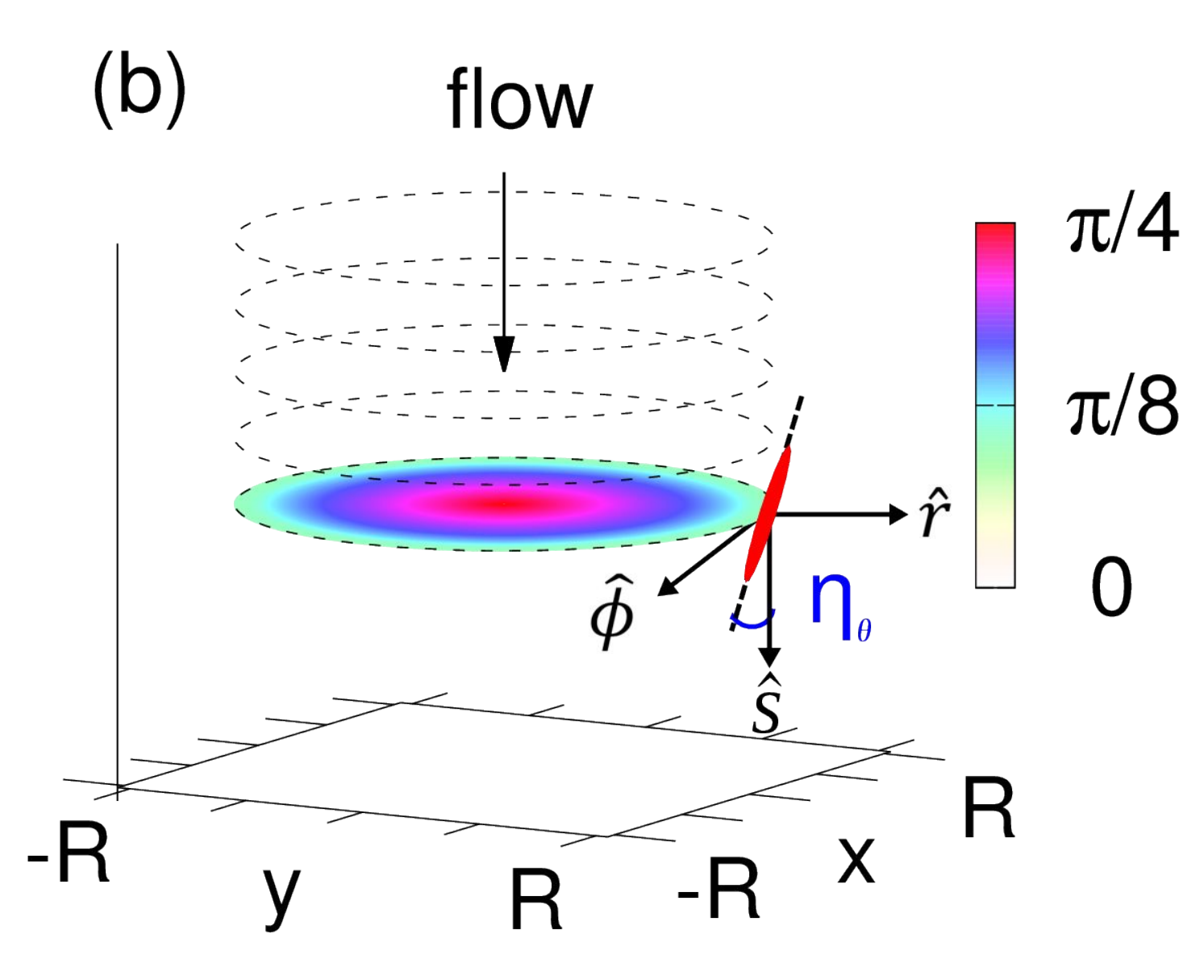}}
 \end{minipage}
\caption{(a) Polar angle $\eta_\theta$ (Eq. \ref{eq:ellipseangles}a) shown in the $xy$-plane and corresponding ellipse located at the outer edge of the nozzle and, (b) $\eta_\theta$ shown in $xyz$-space with arrows indicating the local polar coordinate axis $(r,\phi,s)$ for the ellipse, for $Z_{eq}=37, \beta=0.3$ and $\overline{Wi}_N=2$. The azimuthal angle (Eq. \ref{eq:ellipseangles}b) $\eta_\phi=0$. }
\label{fig:nozangle}
\end{figure}

\begin{figure*}[t]
\begin{minipage}[t]{4.3cm}
 \centerline{\includegraphics[width=4.3cm]{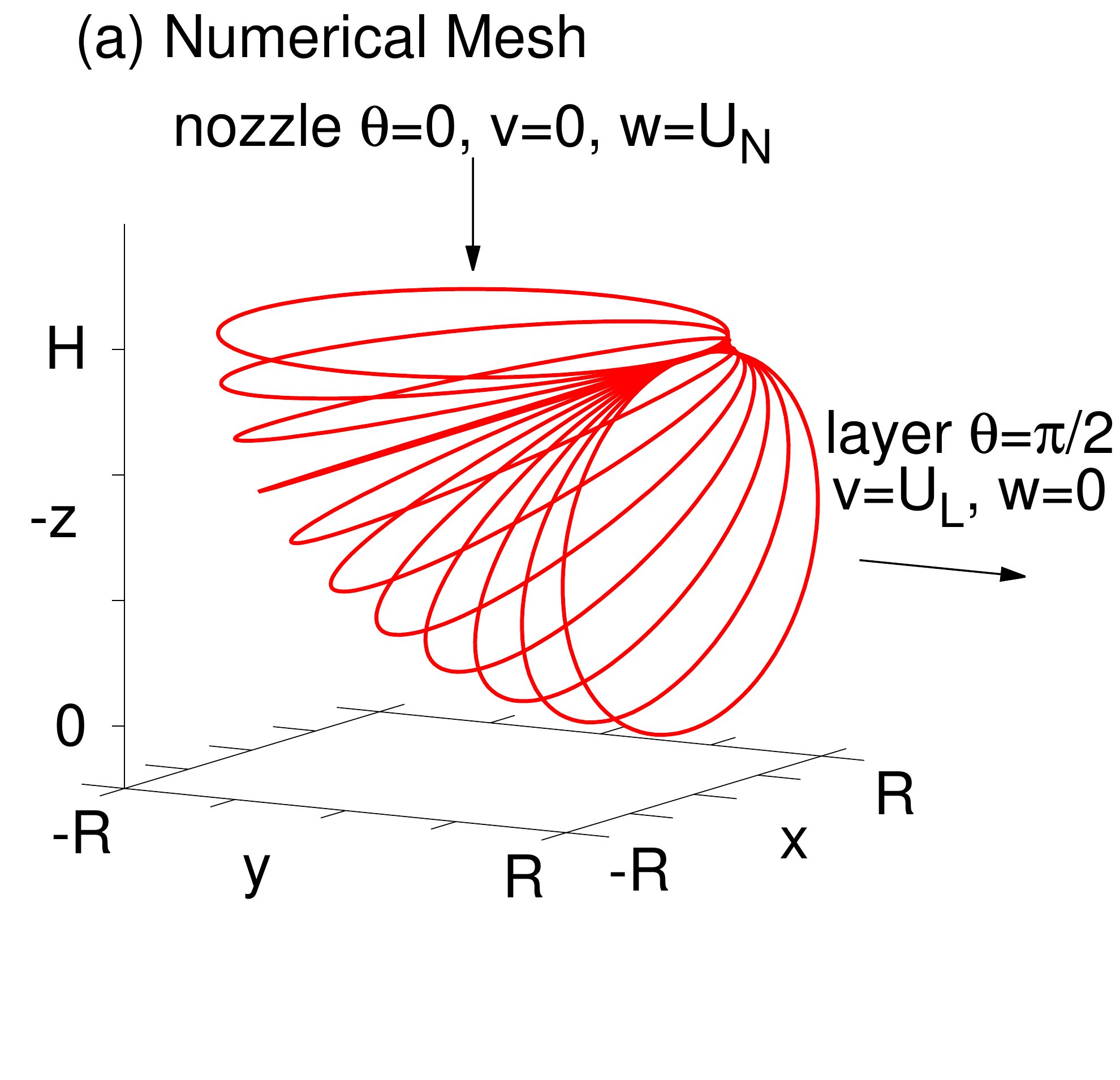}}
\end{minipage}
\begin{minipage}[t]{4.3cm}
\centerline{\includegraphics[width=4.3cm]{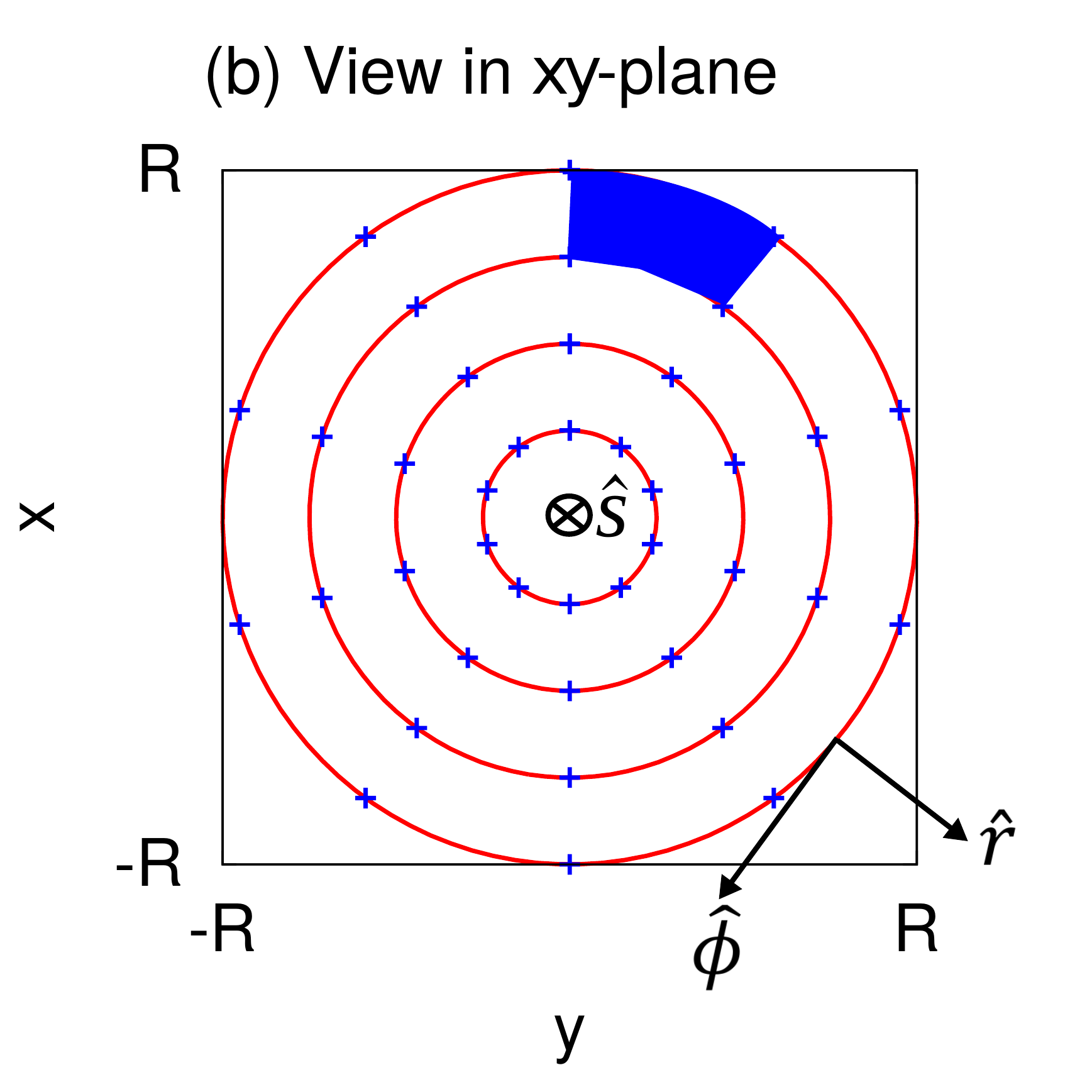}}
\end{minipage}
\begin{minipage}[t]{4.3cm}
\centerline{\includegraphics[width=4.3cm]{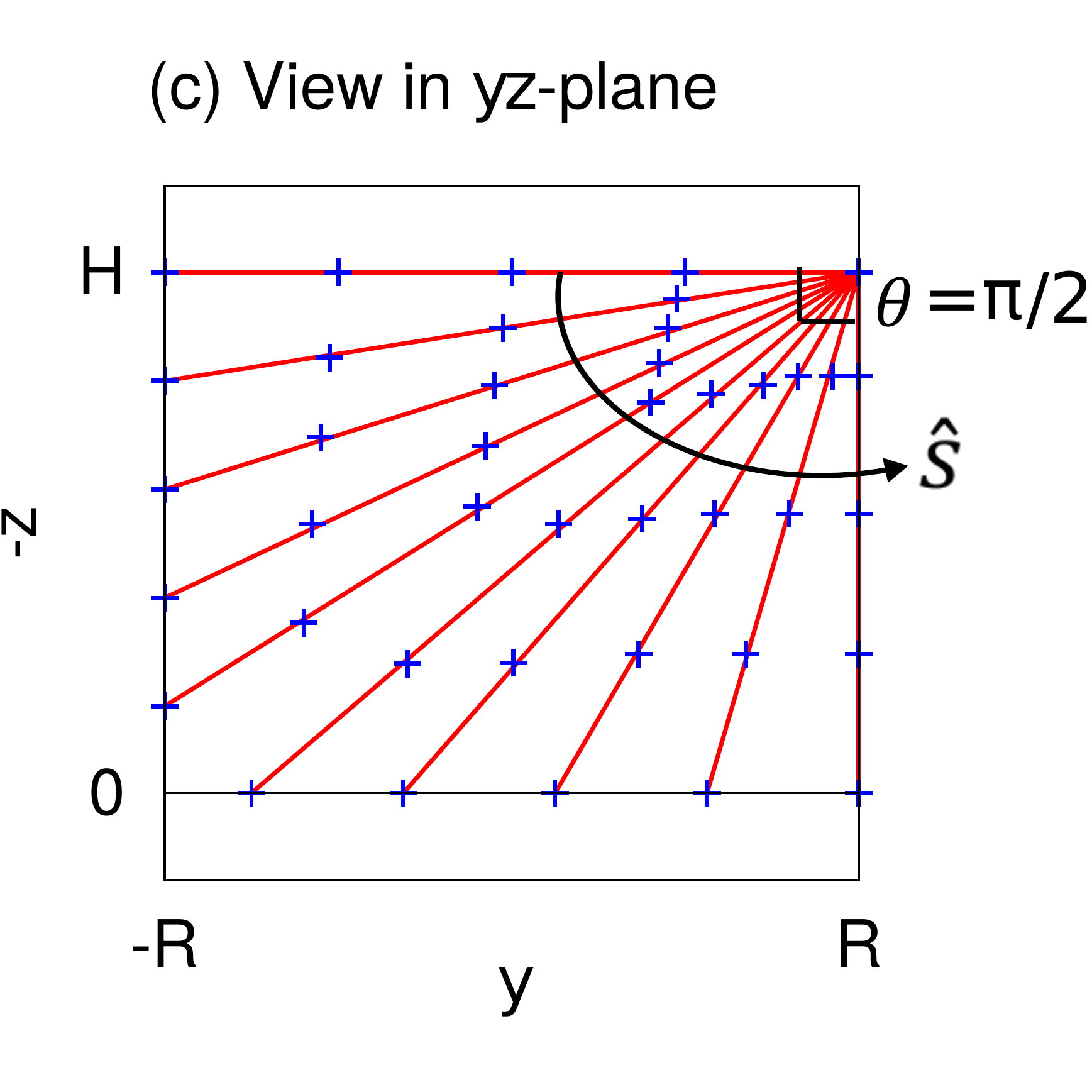}}
\end{minipage}
\begin{minipage}[t]{4.3cm}
\centerline{\includegraphics[width=4.3cm]{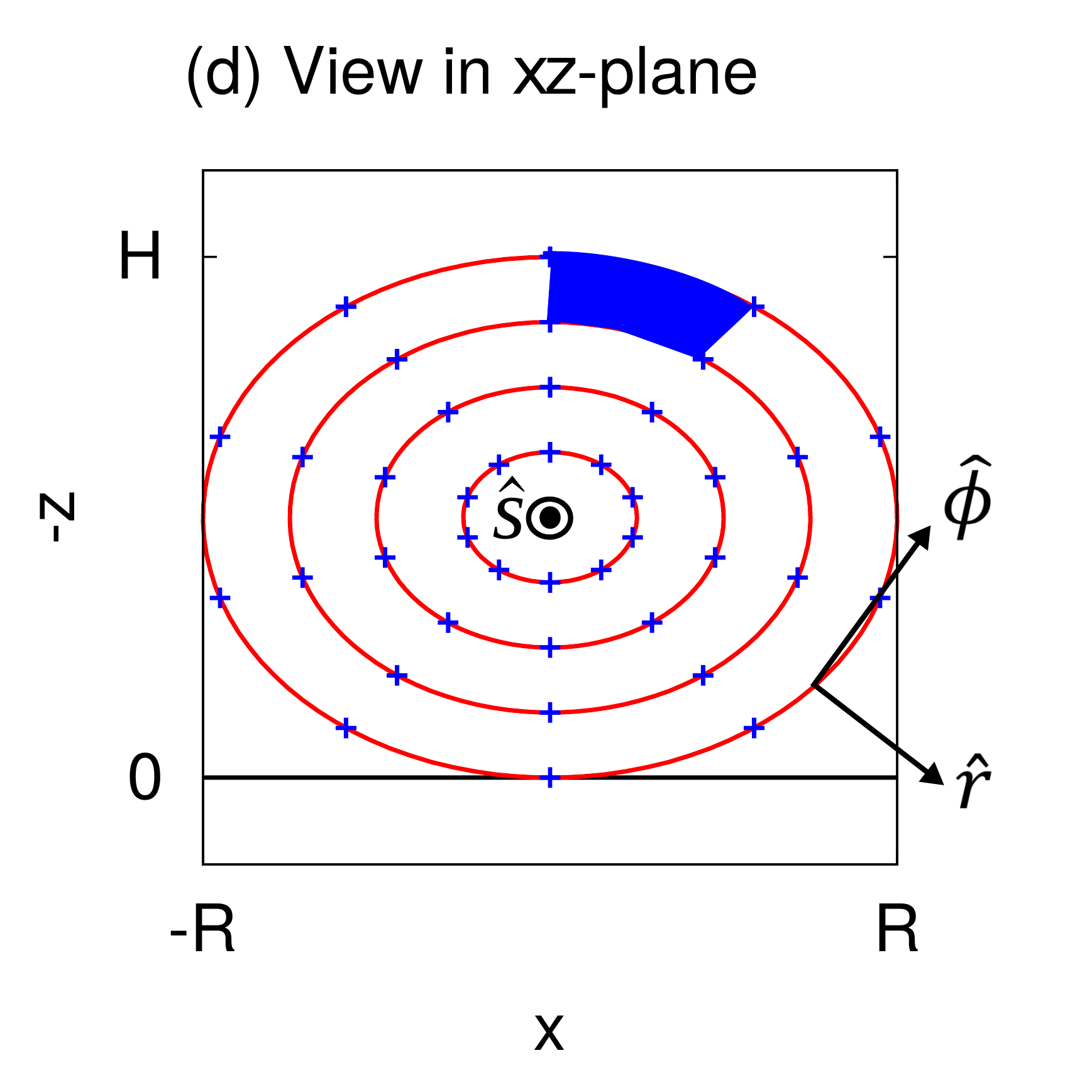}}
\end{minipage}
\caption{(a) Shape of the deposition parametrised by deforming a cylinder into an elliptic-cylinder such that the outside edge of the deposition traces an ellipse. Initial circular cross section ($\theta =0$) has velocity $U_N$ in the $\hat{z}$-direction and final elliptic cross section ($\theta = \pi/2$) has a velocity $U_L$ in the $\hat{y}$-direction. (b) Nozzle view ($\theta=0)$ in $xy$-plane, (c) side view in the $zy$-plane and (d) layer view $(\theta = \pi/2)$ in $xz$-plane; $r$ denotes the radial position on a plane and $\phi$ denotes the angle around a plane. The points indicate individual mesh points and the shaded area represents the area of a mesh element.}
 \label{fig:mesh}
\end{figure*}

Fig. \ref{fig:nozdeform2} shows the contours of the described deformations across the nozzle for the case $\overline{Wi}_N=2$; the axisymmetry should be compared with later results (section \ref{sec:layer}) showing non-axisymmetric deformation after deposition. In particular, Fig. \ref{fig:nozdeform2}a,b highlights the thin boundary layer of disentanglement and stretch at the nozzle wall due to the shear-thinning nature of the flow.  The tensor component $A_{rs}$ parametrises the principle shear deformation of the polymer chain and so Fig. \ref{fig:nozdeform2}c demonstrates the shear stress across the nozzle. Since $A_{s \phi}=  A_{r\phi}=0$ in axisymmetric flow, the shear deformation is solely responsible for the polymer orientation.

This polymer orientation is interpreted as an ellipse with an orientation defined by a polar angle $\eta_\theta$ and azimuthal angle $\eta_\phi$, i.e.
\begin{subequations}
 \label{eq:ellipseangles}
 \begin{align}
  {\bf \hat{e}}_1 \cdot {\bf \hat{s}} &= \cos \eta_\theta, \\
  {\bf \hat{e}}_1 \cdot {\bf \hat{r}} &= - \sin \eta_\theta \cos \eta_\phi , 
 \end{align}
\end{subequations}
where ${\bf \hat{e}}_1$ is the principle eigenvector (corresponding to the largest eigenvalue $\lambda_1$) of the deformation tensor ${\bf A}$.  Fig. \ref{fig:nozangle} shows how the polar angle $\eta_\theta$ decreases near the nozzle wall, demonstrating how the polymer becomes more extended and therefore better aligned with the flow direction in this region. Due to the axisymmetric Poiseulle flow, the azimuthal angle $\eta_\phi$ is zero everywhere and corresponds to ellipses that are tilted `inwards' towards the centre of the nozzle.

\section{Steady-State Deposition Flow}
\label{sec:deposition}

\subsection{Assumptions of Model Deposition Flow}

Due to the small Reynolds number (Re $ \sim 10^{-6}$), the exiting flow quickly assumes a uniform plug-flow velocity profile. The filament shape during deposition is a complicated balance of surface tension, polymer relaxation and complex boundary conditions including the free surface. Although it is known that the material must turn a $90^\text{o}$ bend, the actual deposition shape, corresponding flow field, and temperature profile have yet to be analysed either theoretically or numerically. Rather than solve the full problem, we make the following assumptions. 

\begin{enumerate}
 \item We assume that the temperature is uniform (at $T_N$) during deposition. The extrudate exits the nozzle and reaches the build plate on the time scale
\begin{equation}
\tau_{dep} = H/U_N.
\label{eq:deptime}
\end{equation}
For our model parameters the deposition time is typically of order $\tau_{dep} = 0.005 - 0.03$ s (see Table \ref{tab:speeds}). Upon exit, the material will cool via a combination of convection and radiation. Thus, a non-uniform temperature profile with a cool boundary layer near the free surface, where the layer thickness depends on the print speed, is expected (see Appendix \ref{sec:appendix2}). This cooling will consequently delay polymer relaxation due to the diverging relaxation time (Eq. \ref{eq:shift}). We neglect the effect of this cooling in our model and assume that the temperature of the deposit is uniform. This assumption is roughly compensated by assuming that the polymer does not relax during the deposition stage, as addressed next.


\item  We assume that the deposition occurs sufficiently fast that we can ignore polymer relaxation. For polycarbonate of $Z_{eq}=37$, this requires the deposition time to satisfy
\begin{equation}
\begin{split}
 \tau_{dep} \ll \tau_d^{eq} &= 0.03 \text{ s at } T_N, \\ 
 \tau_{dep} \ll \tau_R^{eq} &= 5.7 \times 10^{-4} \text{ s at } T_N.
 \end{split}
\end{equation}
Although we estimate $\tau_R < \tau_{dep} \lesssim \tau_d$ (see Table \ref{tab:speeds}), a cooling temperature profile as addressed in item 1 may arrest relaxation in the skin layer responsible for welding in a similar way. 
 
\item   We assume that the length scale $z_M$ for which die swell develops is greater than layer thickness $H$; i.e. $z_M >H$. The terminal swell distance downstream of the nozzle exit is \cite{Cloitre:1998}
\begin{equation}
z_M = \tau_{sw} U_N,
\label{eq:swelldistance}
\end{equation}
where $\tau_{sw}$ is the characteristic time scale for the swell diameter to fully develop. This time scale is associated with the relaxation of the first normal stress difference \cite{Allain:1997}, but little known about this relaxation mechanism. Experimentally, $z_M$ is found to be of the order $2R$ and some experiments show $\tau_{sw}$ to depend on the nozzle shear rate \cite{Bird:1987}. For the Rolie-Poly model, the first normal stress difference $N_1 = A_{ss} - A_{rr}$ relaxes on the order of the reptation time $\tau_d$, although for $Wi_R >1$ linear relaxation does not apply. For polycarbonate at print temperature $T_N=250^\text{o}$C, we find that $z_M/ H \approx  1 -10$ in the typical print speed range (see Table \ref{tab:speeds}). Hence, there is probably insufficient time for maximum swell ratio to develop during deposition of the melt onto the build plate. The terminal swell ratio is discussed in Appendix \ref{sec:appendix2}.

\item We assume a smooth ansatz for the shape of the curved filament and demand that the polymeric material undergo affine flow of the elements along streamlines. The effect of changing the curvature of the corner region is discussed in Appendix \ref{sec:appendix2}. 
\end{enumerate}


\begin{figure*}[t]
\begin{minipage}[t]{7cm}
\centerline{\includegraphics[width=7.5cm]{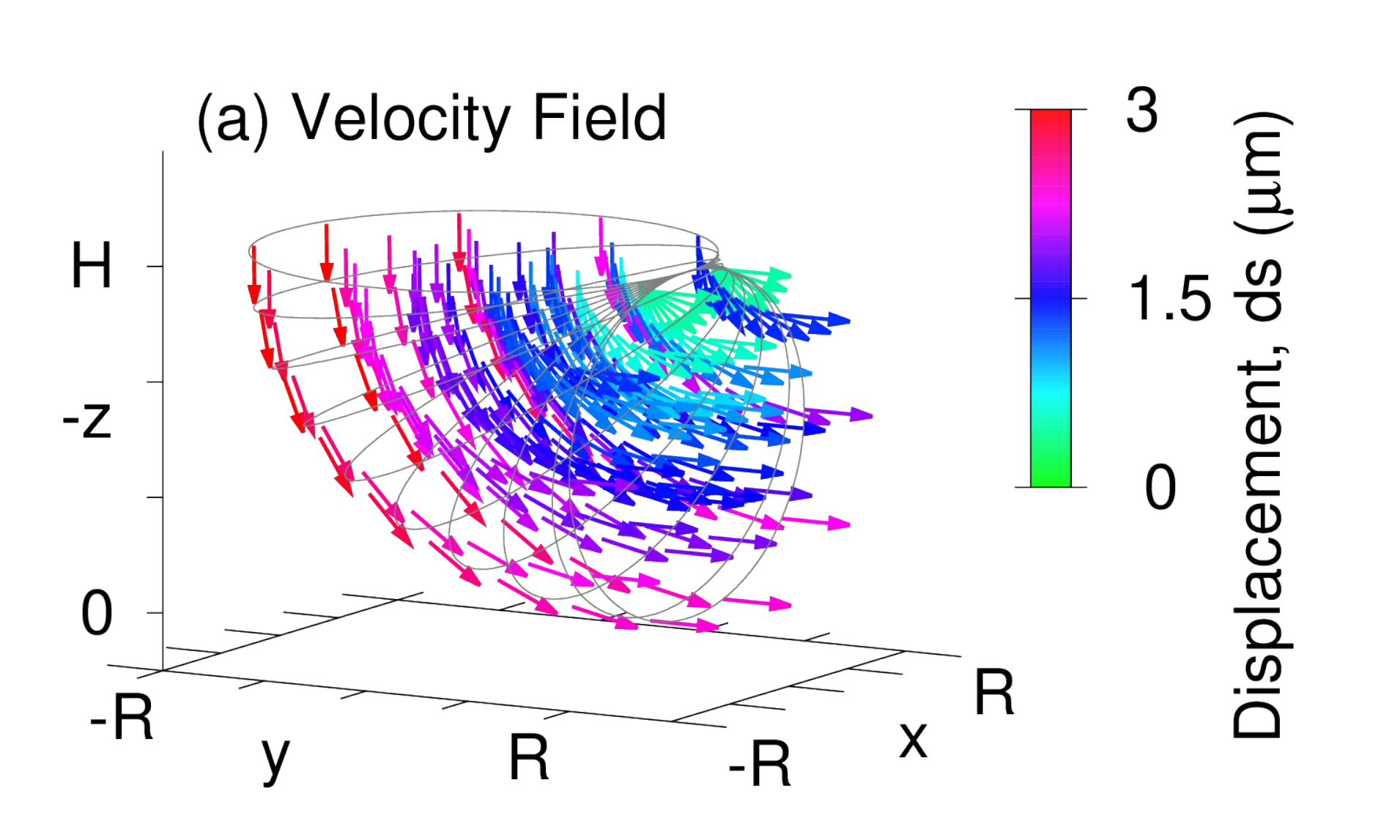}}
\end{minipage}
\begin{minipage}[t]{7cm}
\centerline{\includegraphics[width=7.5cm]{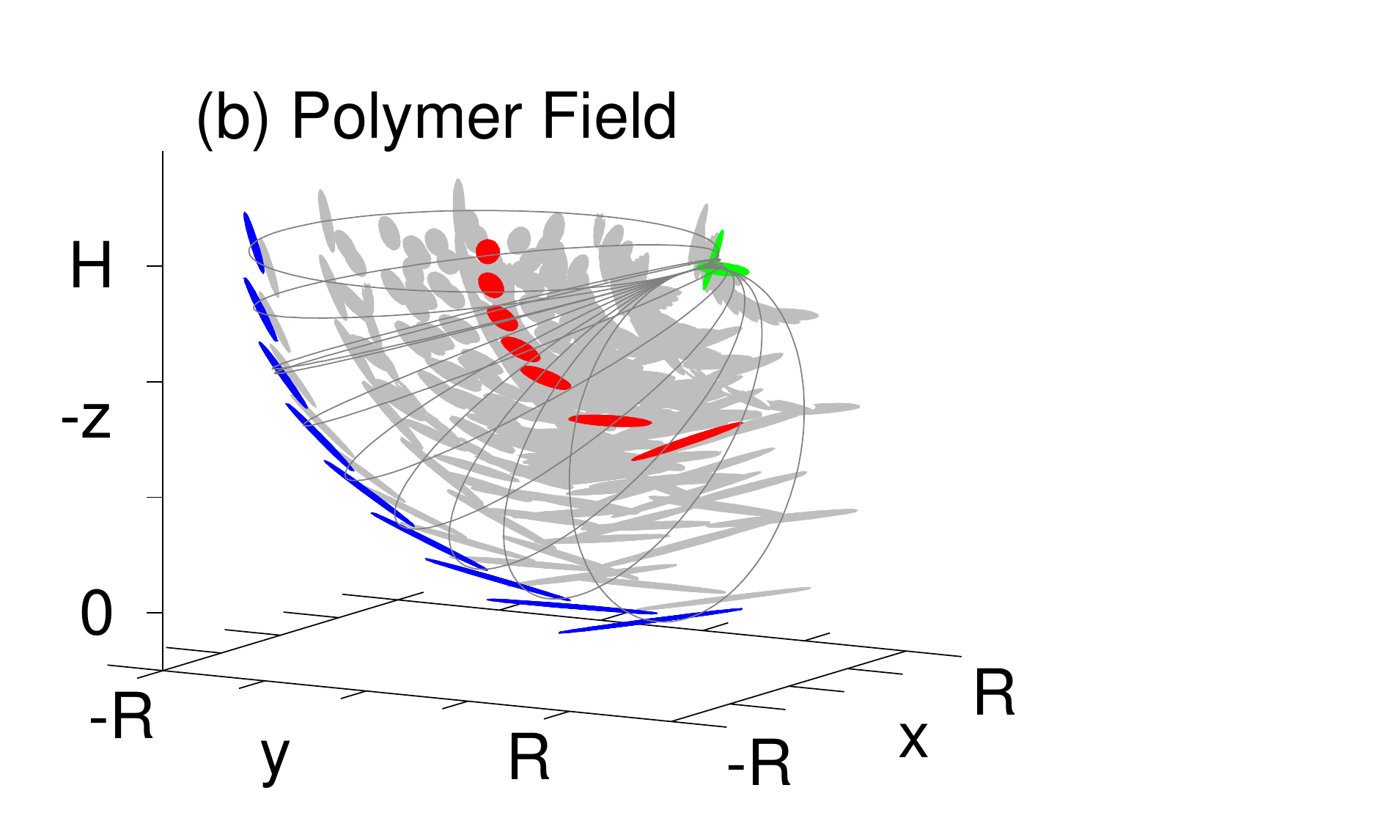}}
\end{minipage}
\begin{minipage}[t]{5cm}
\centerline{\includegraphics[width=5.5cm]{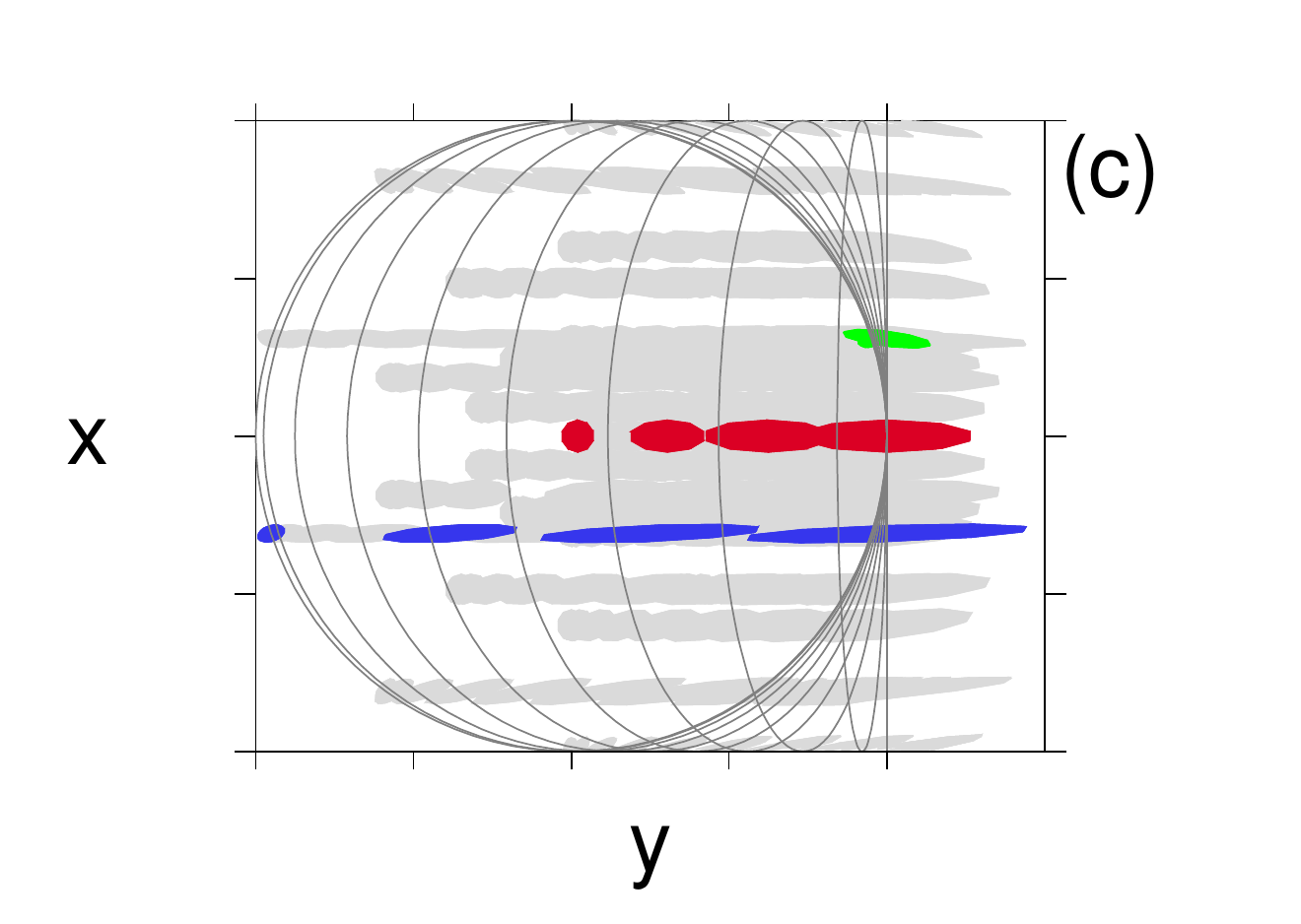}}
\end{minipage}
\begin{minipage}[t]{5cm}
 \centerline{\includegraphics[width=5.5cm]{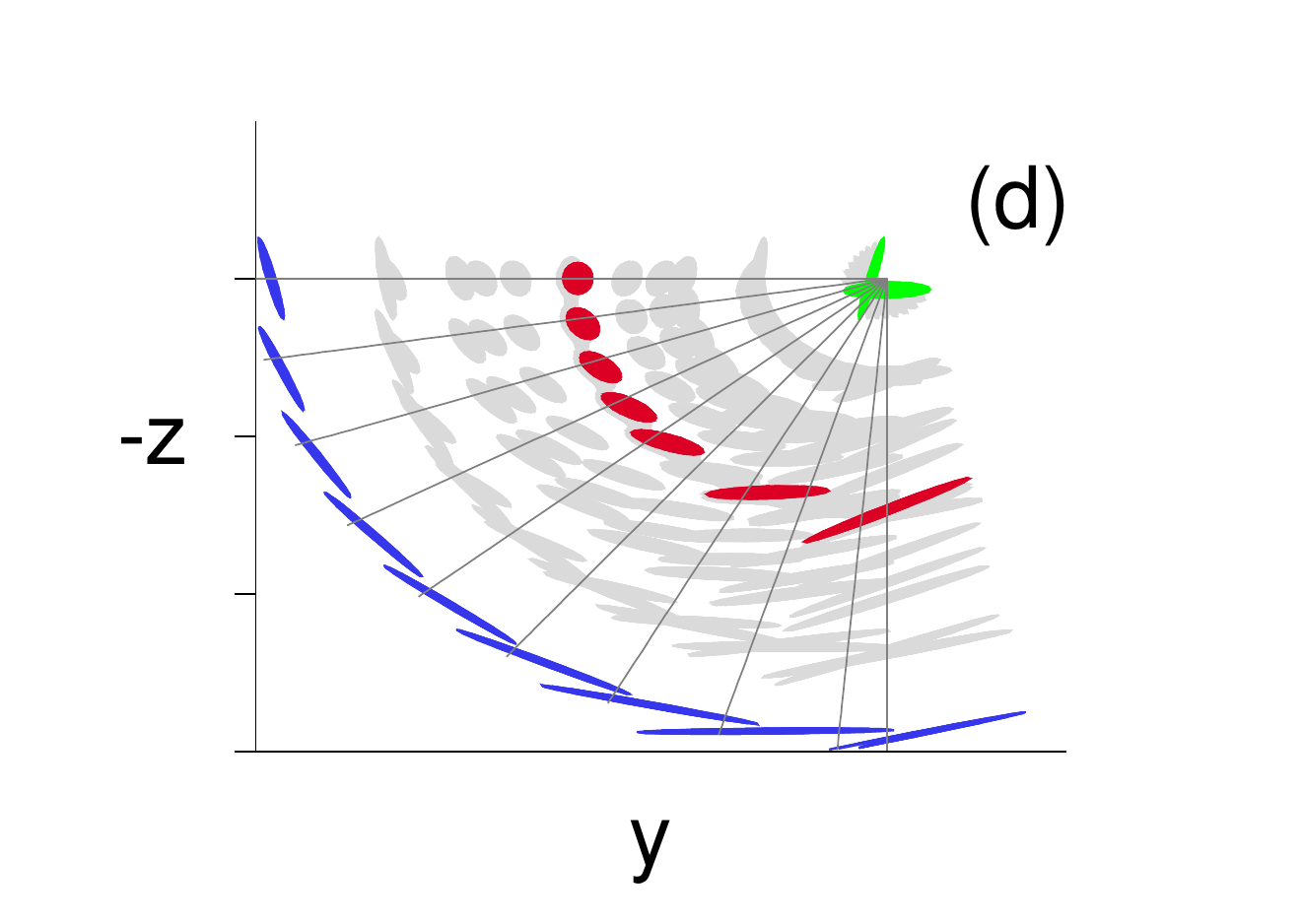}}
\end{minipage}
\begin{minipage}[t]{5cm}
\centerline{\includegraphics[width=5.5cm]{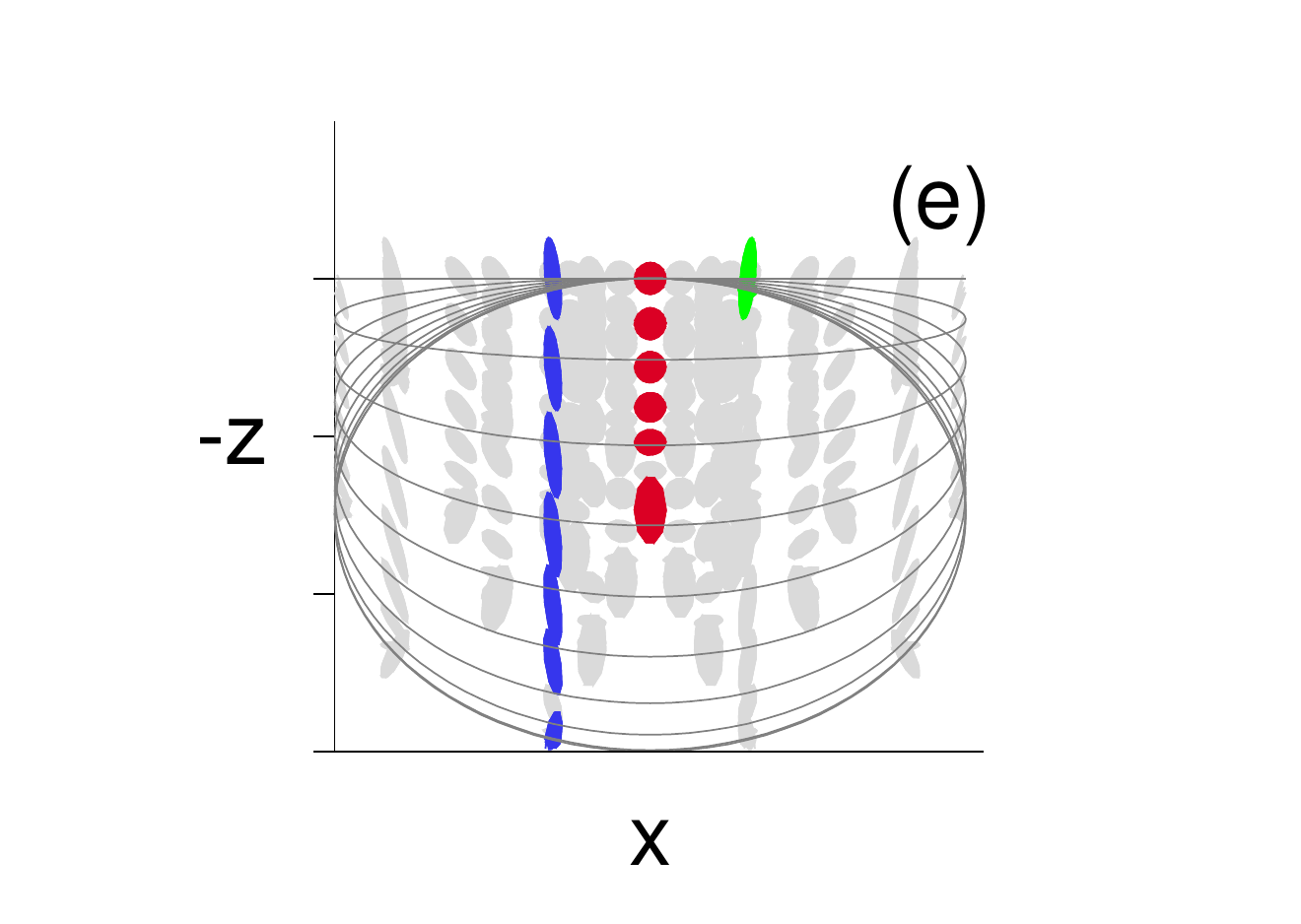}}
\end{minipage}
\caption{(a) Velocity vector field ${\bf u}$ with shading to indicate the displacement $ds$ ($\mu$m). (b) Visualisation of polymer chains as a deformed spheres during deposition process. Colour indicates three locations; top of layer {\em (green)}, middle of layer {\em (red)} bottom of layer {\em (blue)}. Views in (c) the $xy$-plane $(\theta =0)$ looking down the nozzle, (d) the $yz$-plane, showing a side view of the full deposition and (d) $xz$-plane $(\theta=\pi/2)$ looking through the printed layer. Model parameters at $Z{eq}=37, \beta=0.3$ and $\overline{Wi}_N=2$.}
\label{fig:visualdep}
\end{figure*}

\subsection{Deposition Flow and Polymer Deformation}

To parametrise the shape and flow field of the curved filament, we deform a cylinder around a $90^\text{o}$ corner into an elliptic-cylinder, so that the outer edge of the deposition traces an ellipse as shown in Fig. \ref{fig:mesh}. Full details of the mesh generation are given in Appendix \ref{sec:appendix3}. The angle $\theta$ between the nozzle and the fully-deposited layer is in the range $ \theta \in [0,\pi/2]$. In the lab frame, the Cartesian velocity profile ${\bf u}=(0,v,w)$ of the deformation flow is given by 
\begin{equation}
 {\bf u} = U(\theta) \hat{\bf s}(\theta),
\end{equation}
where
\begin{equation}
\hat{\bf s} = \sin\theta \hat{\bf e}_y + \cos \theta \hat{\bf e}_z,
\end{equation}
is the flow direction and $U$ is the magnitude of the velocity. The velocity vector field is shown in Fig. \ref{fig:visualdep}a.

Under the assumption $\tau_{dep} \ll \tau_d, \tau_R$ and assuming no secondary flows, instead of solving the full Navier-Stokes equations the velocity profile is calculated from the flux-conservation condition 
\begin{equation}
 U(\theta) d\mathcal{A}(\theta) = U_N d\mathcal{A}(0),
 \label{eq:velcondition}
\end{equation}
where $d \mathcal{A}$ denotes the area of a cross-section of the deposit at angle $\theta$. This is equivalent to imposing local flux conservation on a single mesh element during deposition (see Fig \ref{fig:mesh}b,c) and is discussed in detail in Appendix \ref{sec:appendix3}. 

Eq. \ref{eq:velcondition} dictates an increase in $U$ to conserve mass during the typical geometric transformation from a circle to an ellipse (Eq. \ref{eq:massconservation}). There is a larger displacement $ds$ towards the outer edge of the deposit to accommodate the $90^\text{o}$ corner (Fig. \ref{fig:visualdep}a); this displacement is given by the arc length
$ ds = r_1 \delta \theta$,
where $r_1$ is the radius measured from the inner corner $(0,R,H)$ and $\delta \theta$ is angle between two cross-sections (see Appendix \ref{sec:appendix3} for further details).

Under this assumption ($\tau_{dep} \ll \tau_d, \tau_R$), the steady-state Rolie-Poly Eq. \ref{eq:Rolie-Poly} is reduced to 
\begin{equation}
 ({\bf u} \cdot \nabla) {\bf A} = {\bf K} \cdot {\bf A} + {\bf A} \cdot {\bf K}^T,
 \label{eq:depRP}
\end{equation}
for velocity gradient tensor
\begin{equation}
 {\bf K} =  \left( \begin{array}{ccc}
0 & 0 & 0 \\
v_x & v_y & v_z \\
w_x &  w_y & w_z \end{array} \right),
\label{eq:depvelocitygradient}
\end{equation}
where the subscripts denote derivatives in the respective directions. In this case, $u_x,u_y,u_z=0$ since there is no change in length in the $\hat{x}$-direction. In this way, the polymer is simply advected with the velocity gradients. Similarly, entanglements are advected via
\begin{equation}
 ({\bf u}\cdot \nabla) \nu = - \beta ({\bf K}:{\bf A}) \nu,
 \label{eq:depnu}
\end{equation}
from Eq. \ref{eq:nu}. 

Eqs. \ref{eq:depRP} and \ref{eq:depnu} are solved using a semi-implicit finite-difference scheme combined with the velocity profile from Eq. \ref{eq:velcondition} and the initial polymer tensor $\bf A$ imposed by the nozzle flow. Full details of the calculation are given in Appendix \ref{sec:appendix3}. For convergence, we require $100$ cross-sections with $200 \times 100$ mesh points on each plane. This corresponds to a mesh element at the outer edge of the printed layer having volume $dx\times dy\times dz = 6 \times3 \times2  \mu\text{m}$ for a nozzle of radius $R=0.2$ mm. 

Fig. \ref{fig:visualdep}b shows an ellipsoidal visualisation of the polymer tensor $\bf A$ during the deposition process for $Z_{eq}=37, \beta=0.3$ and $\overline{Wi}_N=2$. Initially the polymer ellipsoids are directed inwards towards the nozzle centre (as in Fig. \ref{fig:velocity}b). The orientation changes with the flow direction, so that ultimately the polymers are aligned roughly parallel to the printed filament layer. Similar orientations of cellulose fibrils are seen in experiments \citep{Gladman:2016}. Fig. \ref{fig:visualdep}b also shows how the ellipses become more stretched along the outer edge of the deposition due to the increased displacement in this region. Next, we consider the final polymer deformation across the printed cross-section (i.e. for $\theta = \pi/2$).

\begin{figure}[t]
\centerline{\includegraphics[width=9cm]{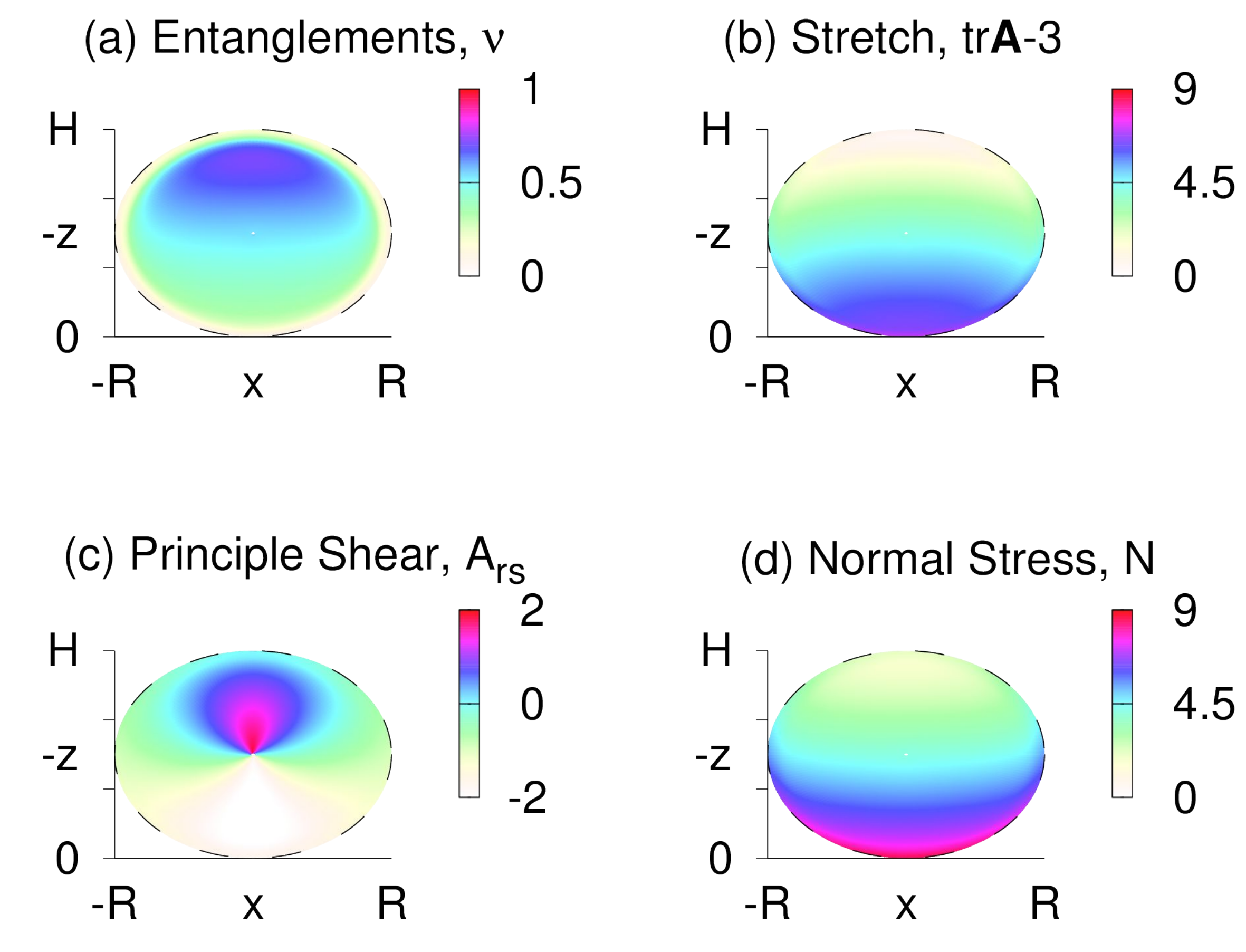}}
\centerline{\includegraphics[width=8.7cm]{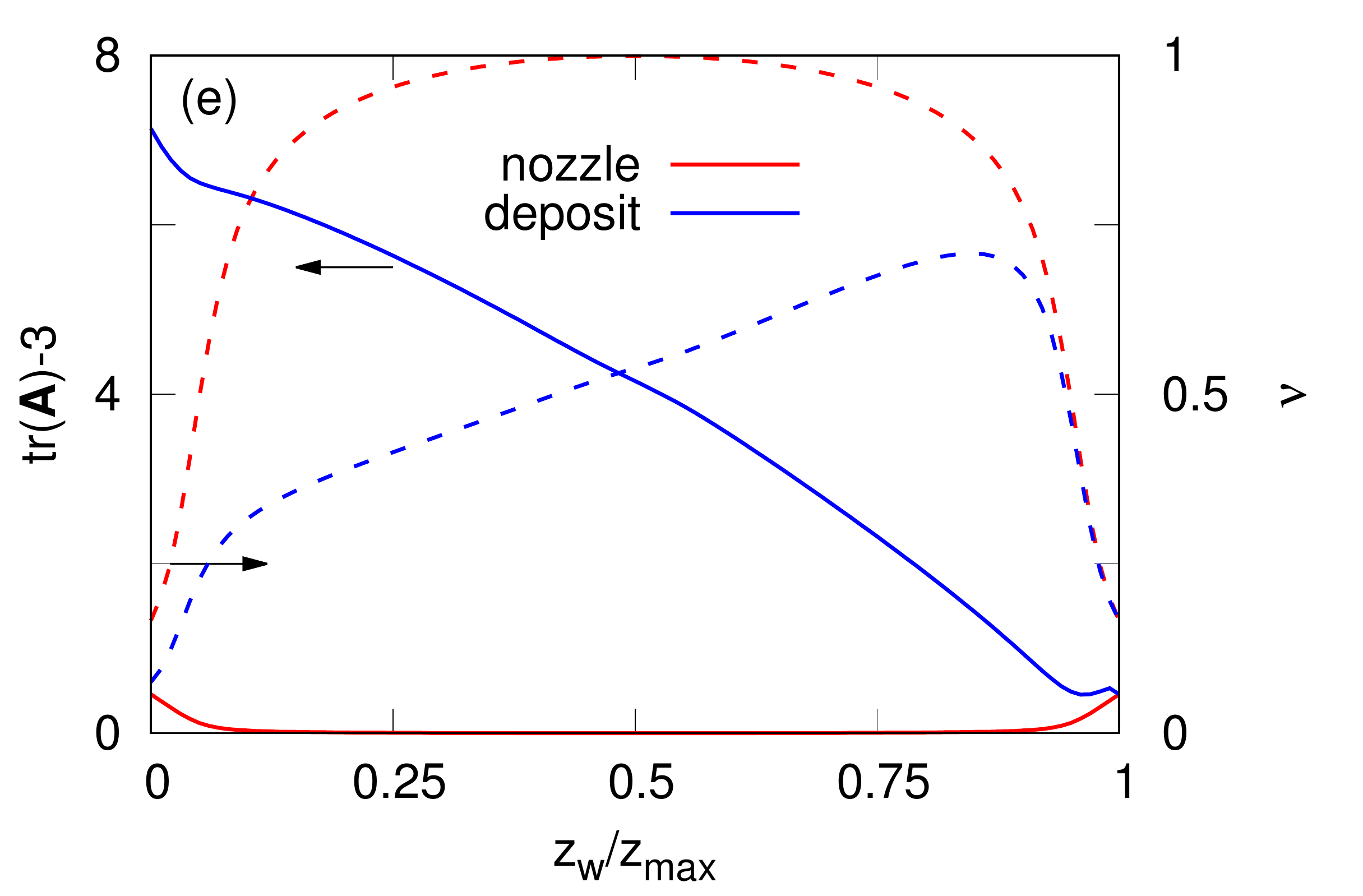}}
\caption{(a-d) Polymer deformation properties after deposition (final plane $\theta = \pi/2$) for $Z_{eq}=37, \beta=0.3$ and $\overline{Wi}_N=2$: (a) Entanglement fraction profile $\nu(r,\phi)$, (b)  tube stretch $\text{tr}{\bf A}(r,\phi)-3$, (c) principle shear deformation $A_{rs}(r,\phi)$ and (d) local normal stress difference $N(r,\phi)$ shown in the $xz$-plane. The fast case induces similar deformation profiles. (e) Quantitative comparison of the stretch {\em (solid line)} and disentanglement {\em (dashed line)} along the $z$-axis ($(x,y)=(0,R)$) induced initially in the nozzle ($z_{max}=2R$) and after deposition across the printed layer ($z_{max}=H$), as a function of distance from weld site $z_w$.}
\label{fig:deposition}
\end{figure}

\begin{figure}[t]
 \centerline{\includegraphics[width=9cm]{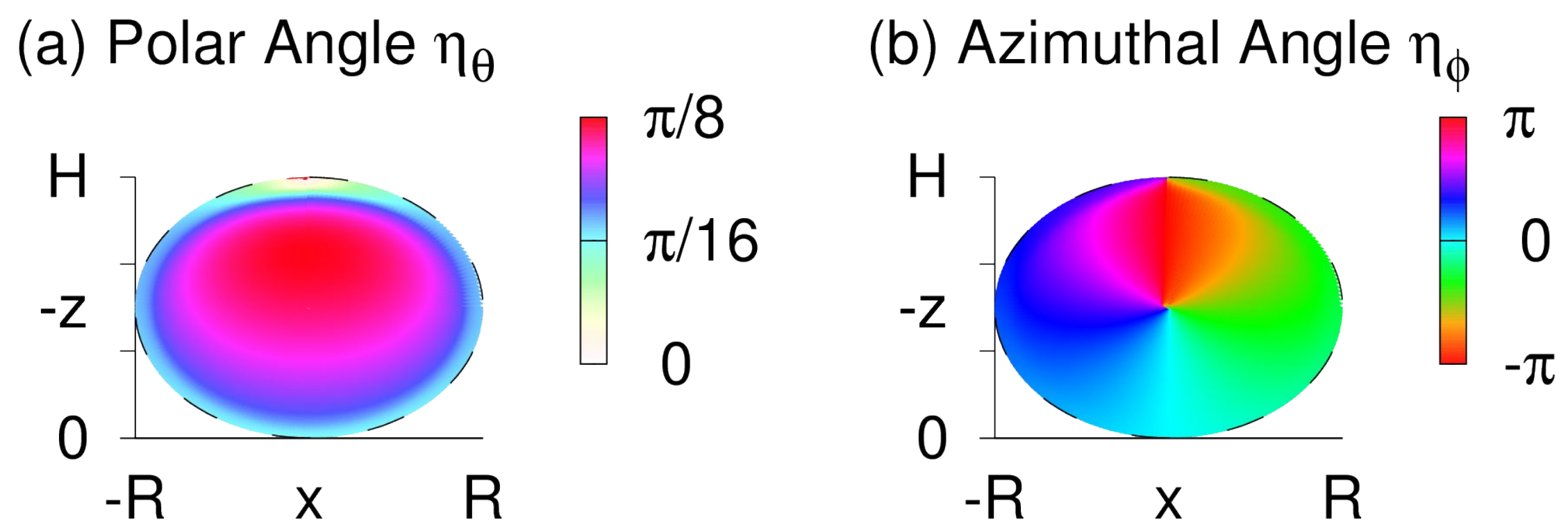}}
 \centerline{\includegraphics[width=8cm]{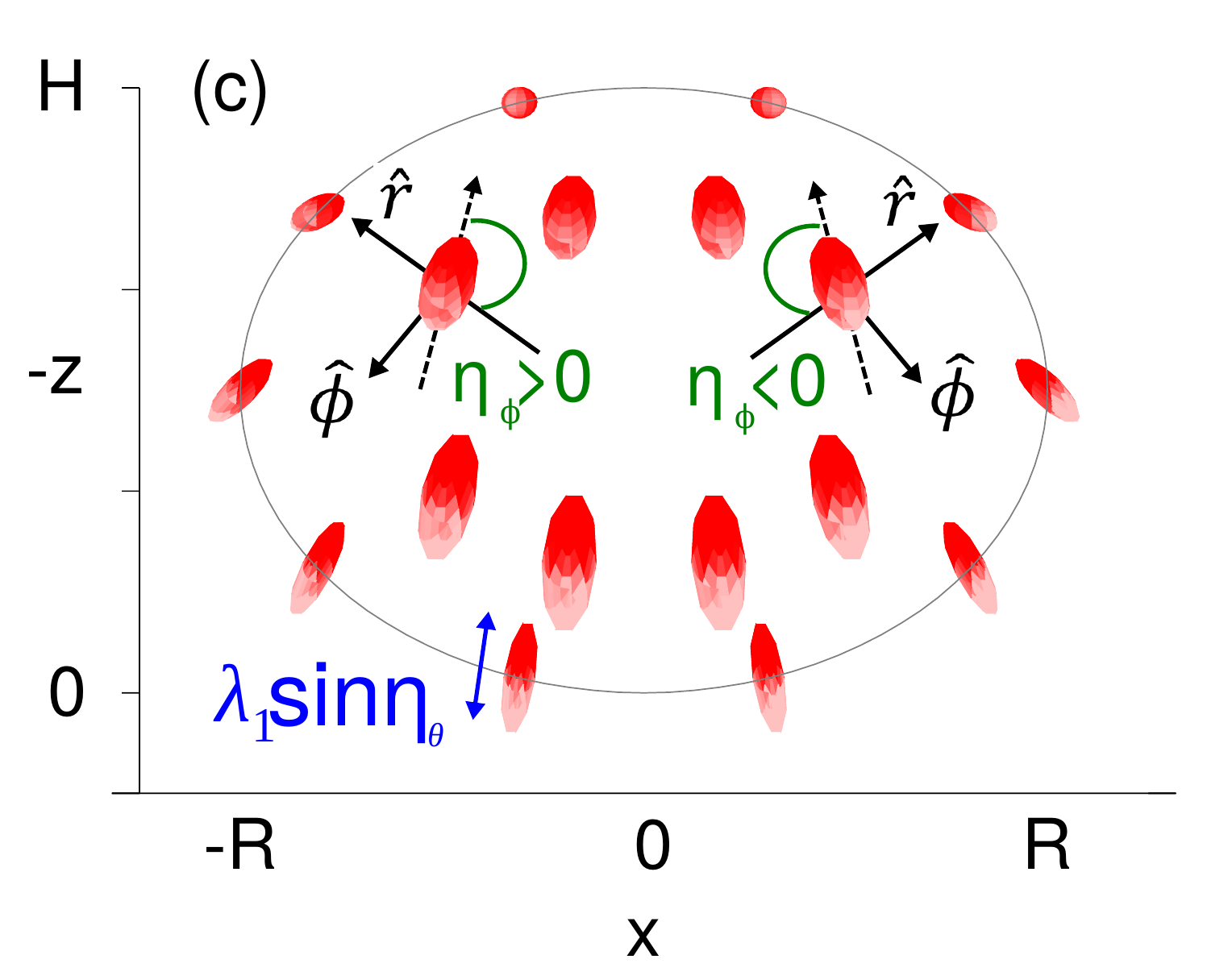}}
\caption{(a) Polar angle $\eta_\theta$ (Eq. \ref{eq:ellipseangles}a) and (b) azimuthal angle $\eta_\phi$ (Eq. \ref{eq:ellipseangles}b) shown in the $xz$-plane; (c) schematic of corresponding ellipses shaded according to $y$-coordinate (red is the front of ellipse) to illustrate angles $\eta_\theta$ (where $\lambda_1$ is the principle eigenvalue) and $\eta_\phi$ (azimuthal rotation from $-\hat{r}$-axis). Arrows indicate the local polar coordinate axis $(r,\phi,s)$. }
\label{fig:angles}
\end{figure}

\section{Results}
\label{sec:layer}


\subsection{Polymer Deformation Across the Printed Filament Cross-section}

Figs. \ref{fig:deposition}a-d show the entanglement fraction $\nu$, the tube stretch $\text{tr}{\bf A}-3$, the principle shear deformation $A_{rs}$ and the normal stress difference $N$ profiles, respectively, across the printed layer for $Z_{eq}=37, \beta=0.3$ and  $\overline{Wi}_N=2$; the deformation profiles are qualitatively similar for $\overline{Wi}_N=13$. In contrast to the nozzle flow (Fig. \ref{fig:nozdeform}), the deformation is no longer axisymmetric and there is a distinct gradient in the polymer microstructure from the top to the bottom of the layer. 

The structure at the top $(z=-H)$ and bottom $(z=0)$ of the layer is of particular interest as welding between adjacent layers in the $\hat{z}$-direction occurs at these sites. The stretch of the free surface due to the curved geometry induces a large deformation along the outer edge of the deposition, so that the polymer microstructure at $z=0$ is highly stretched and oriented (Figs. \ref{fig:deposition}b,c, see Appendix \ref{sec:appendix2} for the effect of changing the curvature). By far, the largest effects occur during this deposition process, with the stretch increasing significantly (by a factor of 3 under the assumption $\tau_{dep} \ll \tau_R$) in the bottom half of the layer. 

Alignment of the polymers in the flow direction, together with the velocity gradient profile, disentangles the polymer melt and $\nu$ is reduced to less than 10\% of the equilibrium entanglement fraction at $z=0$ (Fig. \ref{fig:deposition}a). Although the stretch at $z=-H$ is comparatively smaller, the melt also becomes disentangled in this region (compared to $\nu$ in the nozzle before deposition) due to large velocity gradients. 
Velocity gradients exist primarily due to the material turning 90$^\text{o}$; there is also a secondary contribution due to the transformation of the deposit from a circular to an elliptical shape. 


Due to the anisotropy, the non-axisymmetric components $A_{s \phi}$ and $A_{r \phi}$, corresponding to in-plane tilt and azimuthal shear, respectively, become non-zero and contribute to the total orientation of the polymer. We quantify the effect of these non-axisymmetric components on the polymer orientation by considering the polar and azimuthal angles, $\eta_\theta$ and $\eta_\phi$ (Eq. \ref{eq:ellipseangles}), after deposition. Fig. \ref{fig:angles}a shows that the polar angle $\eta_\theta$ decreases (compared to Fig. \ref{fig:nozangle}), demonstrating how the polymers become more aligned with the flow direction during deposition. The effect of the velocity gradients in the centre of melt is also demonstrated by this decrease in $\eta_\theta$. The corresponding alignment leads to disentanglement at $r=0$, so unlike flow in the nozzle, the entire cross section of the melt becomes disentangled during deposition. The non-zero azimuthal angle $\eta_\phi$ (Fig. \ref{fig:angles}b) signifies how the tilt of the ellipses becomes non-axisymmetric after deposition, with ellipses directed `upwards' away from build plate across the entire layer (Fig. \ref{fig:visualdep}). Ellipses at the top and bottom of the layer have a similar `upwards' tilt, exhibiting a very different orientation to the `inwards' axisymmetric tilt we see in the nozzle.



\begin{figure}[p!]
   \centerline{\includegraphics[width=9cm]{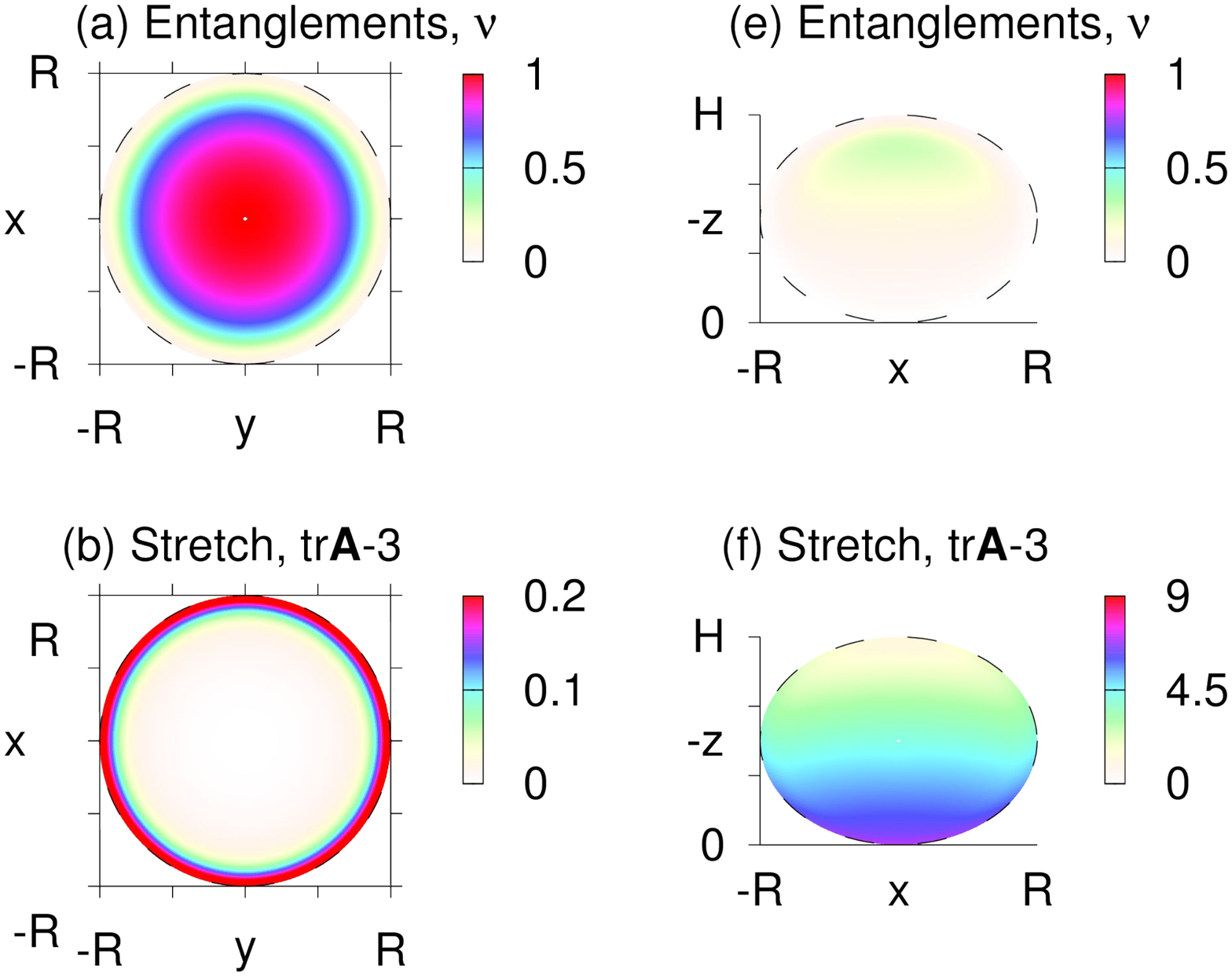}}
   \centerline{\includegraphics[width=9cm]{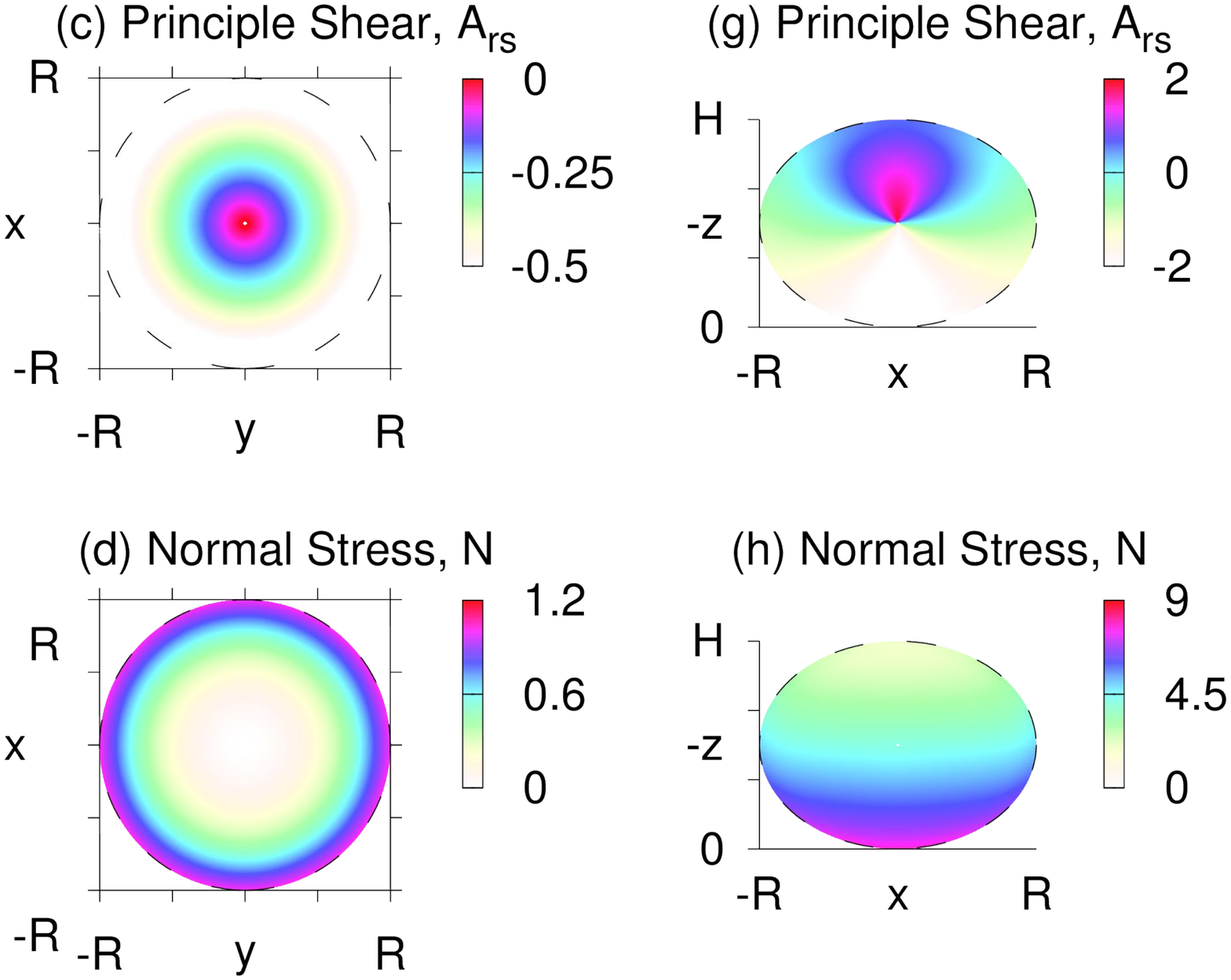}}
   \centerline{\includegraphics[width=8.7cm]{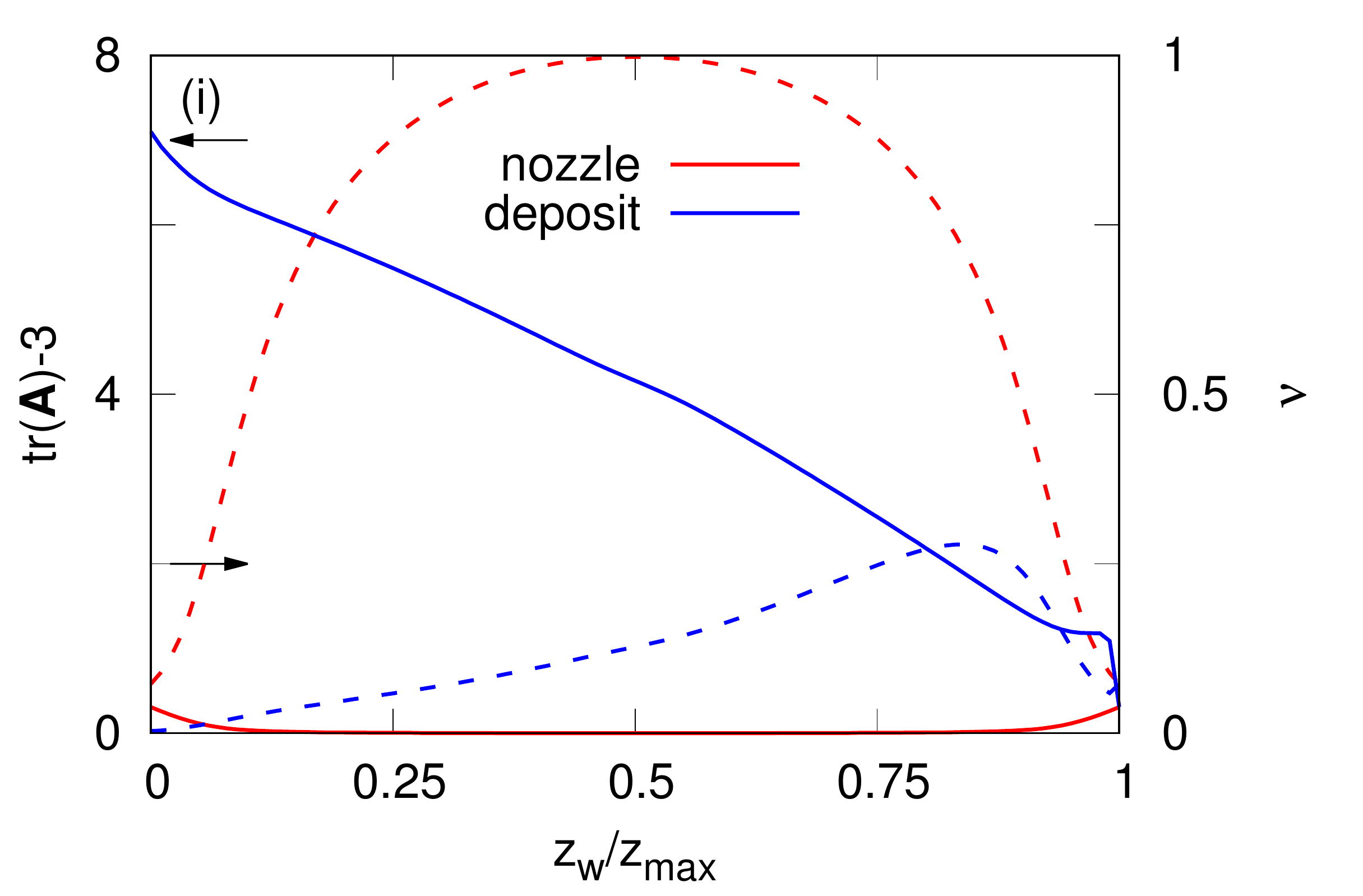}}
 \caption{Polymer deformation properties in the nozzle and after deposition for $Z_{eq}=37$, $\beta=1$ and $\overline{Wi}_N=2$: (a,e) Entanglement fraction profile $\nu(r,\phi)$, (b,f)  tube stretch $\text{tr}{\bf A}(r,\phi)-3$, (c,g) principle shear deformation $A_{rs}(r,\phi)$ and (d,h) local normal stress difference $N(r,\phi)$; (i) Quantitative comparison of the stretch {\em (solid line)} and disentanglement {\em (dashed line)}  along the $z$-axis ($(x,y)=(0,R)$) induced initially in the nozzle ($z_{max}=2R$) and after deposition across the printed layer ($z_{max}=H$), as a function of distance from weld site $z_w$. }
\label{fig:beta0.3}
\end{figure}

\subsection{Effect of CCR parameter on Disentanglement}
\label{sec:CCR}

For comparison with Figs. \ref{fig:nozdeform2} and \ref{fig:deposition} with $\beta=0.3$, Fig. \ref{fig:beta0.3} shows the deformation imposed by the nozzle flow and during deposition for CCR parameter $\beta =1$. Since the melt is less shear-thinning in the case $\beta=1$, there is a larger boundary layer of disentanglement in the nozzle (compared to Fig. \ref{fig:nozdeform2}) and the disentanglement induced by the deposition process is much more extreme (compared to Fig. \ref{fig:deposition}). The deformation of $\bf A$ imposed during deposition is equivalent for $\beta = 1$ and 0.3 due to the assumption $\tau_{dep} \ll \tau_d, \tau_R$ (Eq. \ref{eq:depRP}).

\subsection{Effect of Shear Rate on Disentanglement}

\begin{figure}[t!]
 \centerline{\includegraphics[width=8cm]{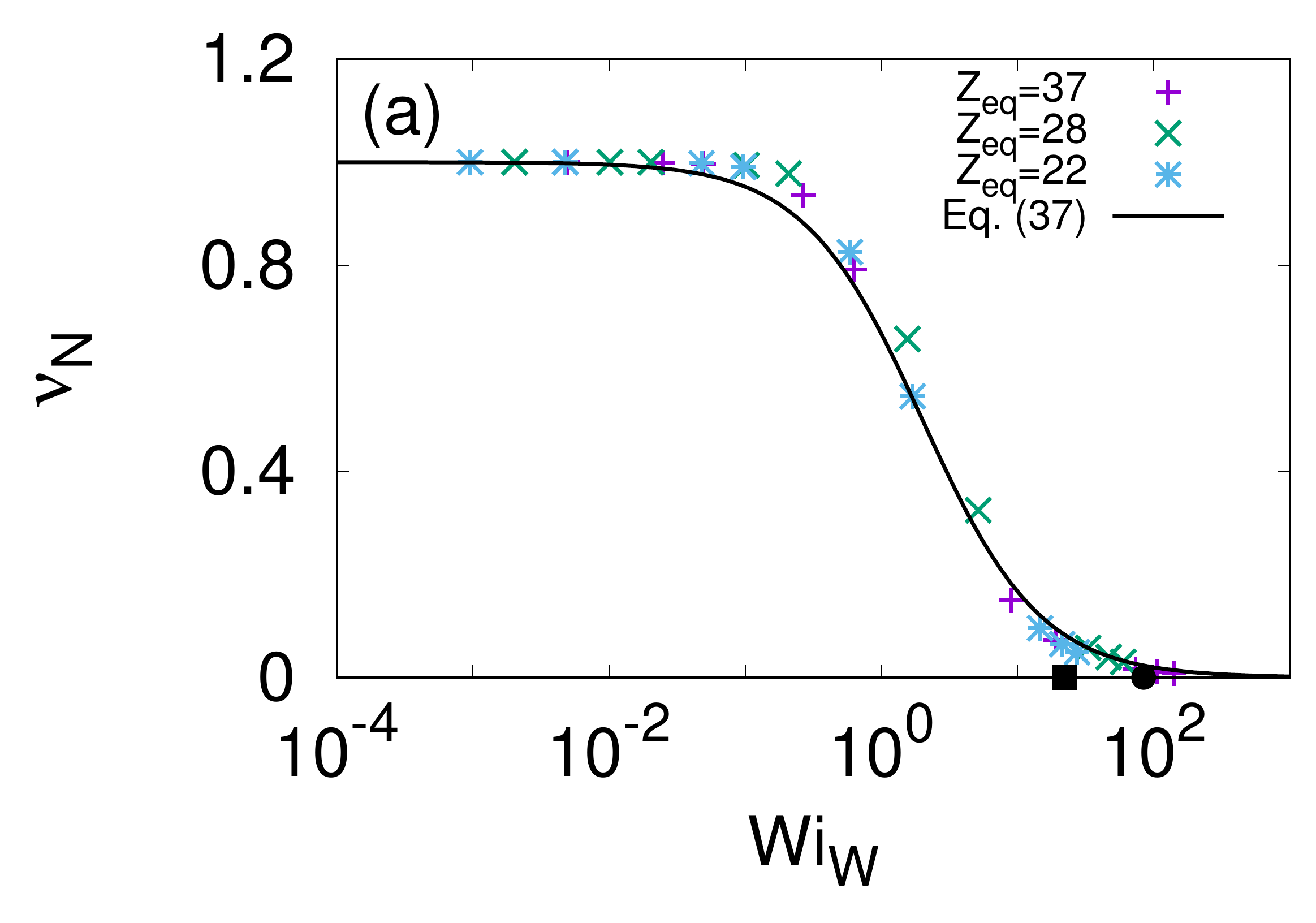}}
 \centerline{\includegraphics[width=8cm]{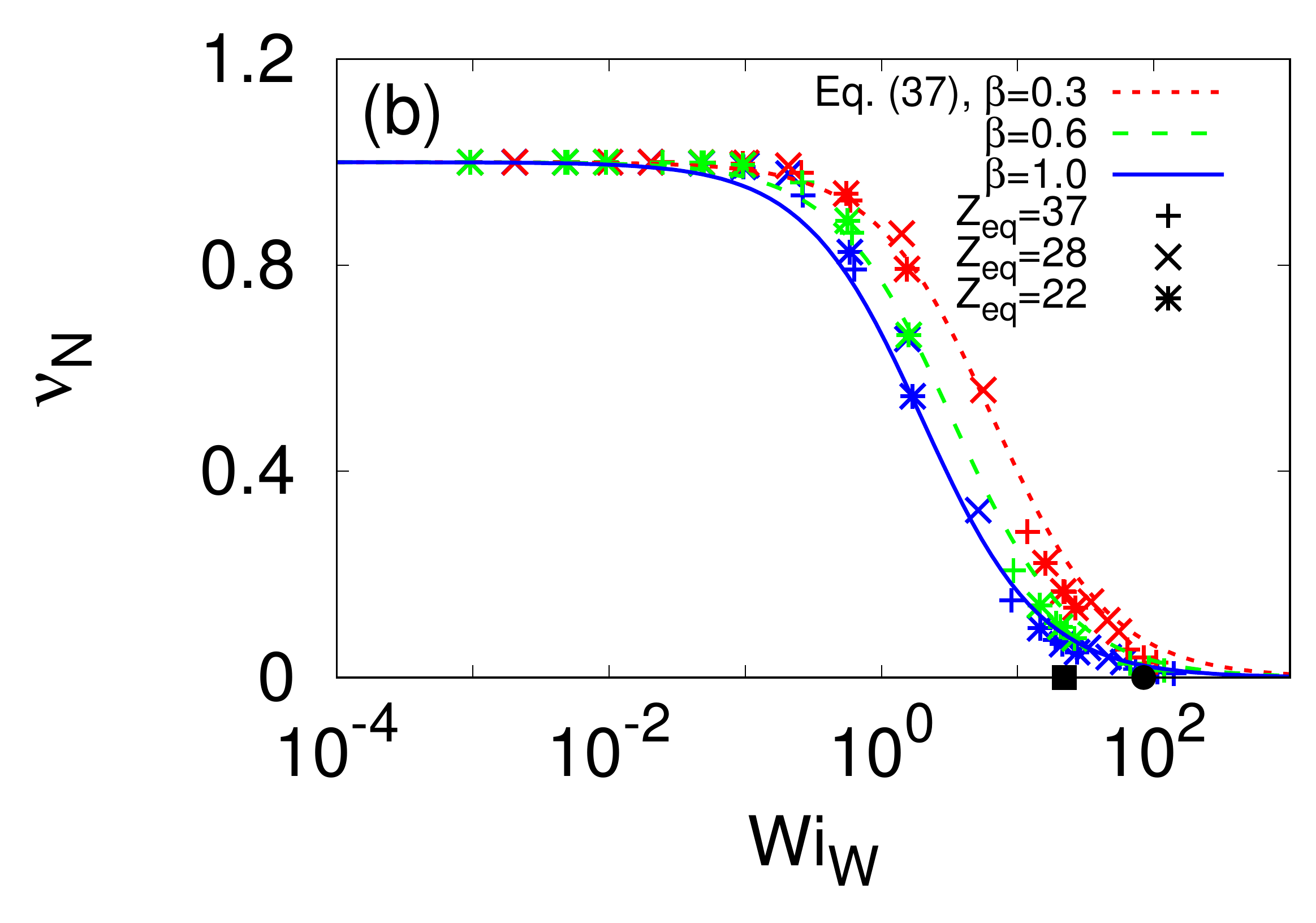}}
 \centerline{\includegraphics[width=8cm]{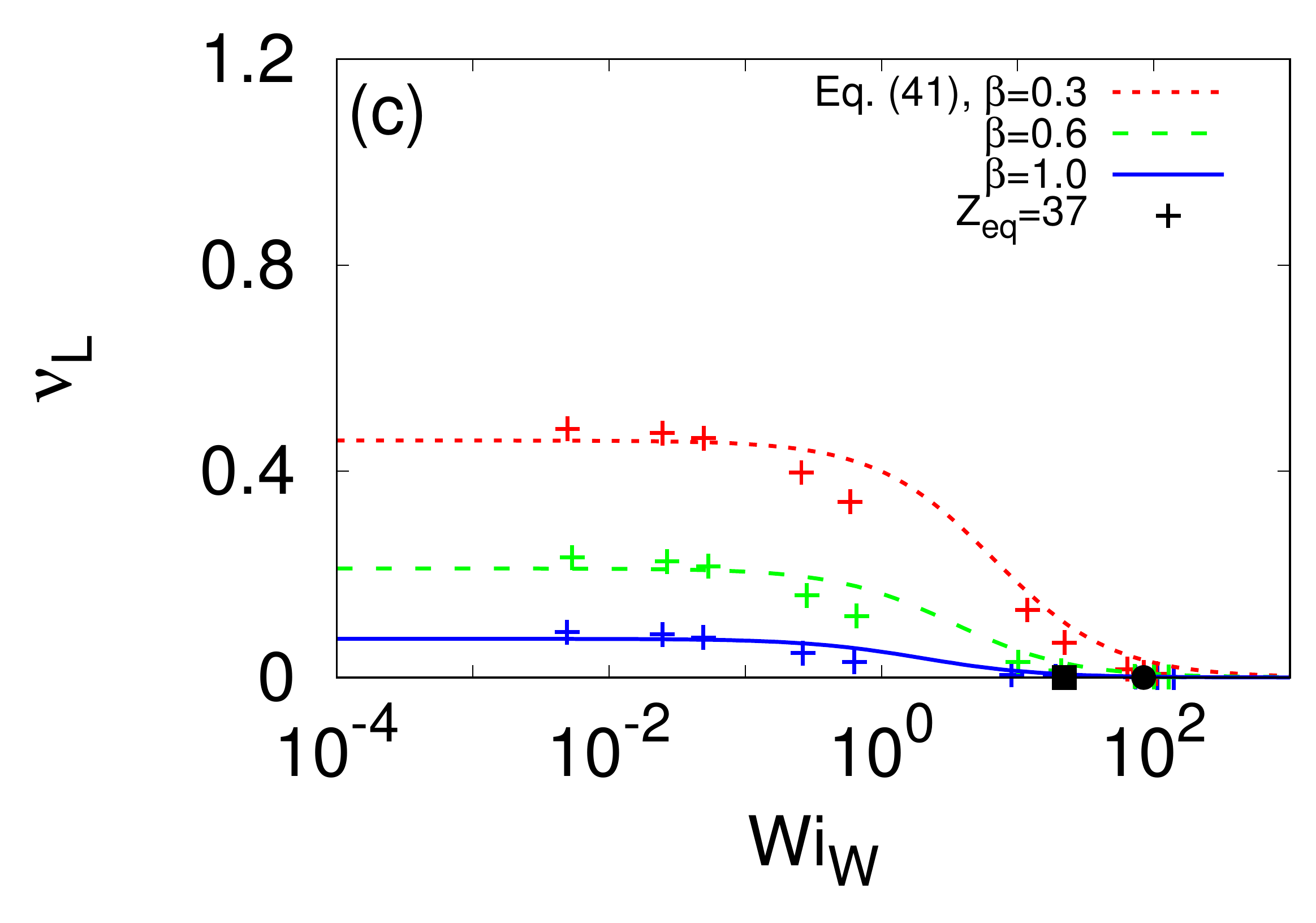}}
\caption{Degree of disentanglement $\nu$ for a range of Weissenberg numbers $Wi_W$ (Eq. \ref{eq:WiW}): (a)  $\nu_N$ at the nozzle wall before deposition for $\beta=1$ and equilibrium entanglement numbers $Z_{eq}$=22 (*), 27 (x) and 37 (+), (b) $\nu_N$ before deposition for a range of CCR parameters $\beta$ and (c) $\nu_L$ at the weld site $z=0$ after deposition for $\beta=1, 0.6$ and 0.3. Theory lines given by Eqs. (\ref{eq:nuN}) and (\ref{eq:nudeptheory}). The black square marks the slow case $\overline{Wi}_N=2$  and the black circle marks the fast case $\overline{Wi}_N=13$. }
\label{fig:disentanglement}
\end{figure}

Here we consider how the predicted disentanglement varies with the equilibrium entanglement number $Z_{eq}$ (equivalent to changing $M_w$) and the local Weissenberg number calculated at the nozzle wall $Wi_W$ (Eq. \ref{eq:WiW}). 

Fig. \ref{fig:disentanglement}a shows the entanglement fraction at the nozzle wall (prior to deposition), $\nu_N$, for $\beta=1$ and three molecular weights. 
From Eq. \ref{eq:nusteady}, $\nu_N$ is given by
\begin{equation}
\nu_N   = \frac{1}{1 + \beta A_{rs} Wi_W},
 \label{eq:nuN}
\end{equation}
and agrees quantitatively with the calculated degree of disentanglement at the nozzle wall for $A_{rs}=0.5$, although $A_{rs}$ is not independent of $Wi_W$. The disentanglement fraction does not depend on $Z_{eq}$.

For $Wi_{W} > 1$, $\nu_N$ is reduced to less than 20\% of the equilibrium entanglement fraction and reducing the CCR parameter slightly inhibits disentanglement at the nozzle wall (Fig. \ref{fig:disentanglement}b). We find  nearly 100\% disentanglement at $Wi_W=100$. For comparison, a 50\% entanglement loss is found for $Wi=100$ in the molecular simulations of simple shear flow by Baig {\em et al.} \cite{Baig:2010}. This is well represented by the theory of Ianniruberto \& Marrucci for $Z_{eq}=14, \beta =0.15$ \cite{Ianniruberto:2014}, which employs the Doi-Edwards tensor rather than the Rolie-Poly model. For our model parameters, $\beta=0.15$ gives a non-monotonic constitutive curve (see Appendix \ref{sec:appendix1}).
 
Fig. \ref{fig:disentanglement}c shows the entanglement fraction $\nu_L$ at weld site $z=0$ after deposition. Again this process is independent of $Z_{eq}$. For the case $\beta =1$, the idealised deposition process imposed by the model removes almost all entanglements from the melt at the weld site. For smaller values of the CCR parameter $\beta$ the disentanglement process is significantly less severe for moderate Weissenberg numbers, and the melt only becomes fully disentangled for $Wi_W \ge 100$.

By spatially advecting the entanglements through the deposition and assuming $\tau_{dep} \ll \tau_d$, Eq. \ref{eq:depnu} leads to
\begin{equation}
 U \frac{\partial \nu}{\partial s} \sim - \beta ({\bf K}:{\bf A})\nu,
\end{equation}
for flow direction $\hat{\bf s}$. Since the deformation is dominated by the extension induced by stretching the fluid elements around the corner, we assume
\begin{equation}
 {\bf K}: {\bf  A} \approx \frac{\partial U}{\partial s} A_{ss},
\end{equation}
which gives
\begin{equation}
 \frac{1}{\nu} \frac{ \partial \nu}{\partial s} \sim -  \frac{\beta A_{ss}}{U} \frac{ \partial U}{\partial s}.
\end{equation}
Integrating yields
\begin{equation}
 \nu_L \sim \nu_N \left( \frac{U_L}{U_N} \right)^{-\beta A_{ss}},
 \label{eq:nudeptheory}
\end{equation}
 where $\nu_N$ is given by Eq. \ref{eq:nuN}. Thus, disentanglement depends on the geometry, which determines the ratio $U_L/U_N$ (Eq. \ref{eq:massconservation}), the total stretch imposed and the CCR parameter $\beta$. Eq. \ref{eq:nudeptheory} fits the data well for $\beta=1.0, 0.6$ and 0.3, and $A_{ss}=9$ (Fig. \ref{fig:disentanglement}c),  although $A_{ss}$ is not independent of $Wi_W$.

\section{Discussion}

\subsection{Model Summary and Limitations}

We have developed a model of the fused-filament-fabrication process and tested the effect of changing print speed, entanglement number $Z_{eq}$ and CCR parameter $\beta$ on the degree of polymer deformation and disentanglement during extrusion. We have used Bisphenol A Polycarbonate as an example of a typical amorphous polymer used for FFF. We model the nozzle flow as axisymmetric, steady-state pipe flow. The nozzle flow can stretch and orient the polymer near the nozzle wall, which consequently disentangles the melt via convective constraint release.

Since the material must melt before being deposited, practically the upper speed limit for printing is restricted by thermal diffusion in the nozzle. In the model, we assume a uniform temperature profile across the nozzle radius. For polycarbonate, it is estimated to take $\sim 7$ s to achieve $T_N$ across the nozzle radius via thermal diffusivity (see Appendix \ref{sec:appendix2}). By comparing to the residence time in the heated nozzle section, this leads to an upper flow rate limit of $\sim 3 \times 10^{-6}$ kg/s for our model, although faster rates are often used (e.g. $9\times10^{-6}$ k/s for $U_L=100$ mm/s).  Arguably only the outer side of the filament must be melted to ensure welding.
Moreover, fluorescence-based measurements during polymer extrusion report temperature gradients of up to $5^\text{o}$C/mm between the centre of the nozzle and the wall due to shear heating effects \cite{Migler:1998}. A more detailed model is required to capture the effects of an inhomogeneous temperature profile in the nozzle.

After exiting the nozzle, the extrudate deforms to make a 90$^\text{o}$ turn and is deposited into a elliptical-shaped layer. Rather than calculate the full fluid mechanics, we have calculated the steady-state flow by assuming flux conservation and a uniform temperature profile. However, we estimate that a cool boundary layer with thickness $\sim 0.1$ mm will develop during deposition (see Appendix \ref{sec:appendix2}); a more detailed model is required to capture these complex cooling dynamics upon exiting the nozzle. We also neglect polymer relaxation during deposition. The assumption that reptation is slow compared to the deposition time ($\tau_d > \tau_{dep}$),  yields the validity condition 
\begin{equation}
\overline{Wi}_N > \frac{H}{R},
\end{equation}
which for our model parameters leads to $\overline{Wi}_N > 1.5$. For smaller Weissenberg numbers where polymer relaxation must be considered, the flow may not be in steady state during deposition.

Due to the corner flow geometry and the transition from a circular to an elliptical shape, the polymer deformation is affected primarily by the deposition flow rather than the nozzle flow, with polymer stretch becoming significant in the bottom half of the layer. During deposition, the polymer tensor ${\bf A}$ is deformed further from equilibrium, non-axisymmetric configurations become non-zero and the structure of the weld region is highly stretched, oriented and partially disentangled. The degree of disentanglement at the weld site $(z=0)$ depends on both the shear rate in the nozzle and the CCR parameter $\beta$. Faster printing imposes a greater deformation and disentangles the melt further during the extrusion process. For $Wi_W> 100$, the weld site becomes fully disentangled for all $\beta$.





Understanding the polymer behaviour during extrusion in terms of the material properties, print speed and nozzle geometry is key to characterising the strength of the weld between printed filaments. After deposition, the printed melt will rapidly cool towards the glass transition. The way in which the deformation relaxes as a function of temperature governs the diffusive behaviour at the weld and is therefore key to understanding the ultimate welding characteristics such as weld thickness, structure and entanglement. The effect of this polymer deformation on welding behaviour will be discussed elsewhere.

\subsection{Outlook}

The molecular CCR mechanism is key to understanding polymer behaviour in highly non-linear flows such as the FFF technique for additive manufacturing. CCR was first added to the original Doi-Edwards tube model by Ianniruberto \& Marrucci \cite{Ianniruberto:1996, Marrucci:1996}. The recent GLaMM model \cite{Graham:2003} refines the tube theory further to include the effects of CCR on the chain stretch. CCR in the tube model has now been revisited to account for flow-induced changes in the entanglement density \cite{Ianniruberto:2014, Ianniruberto:2015}. 

In this paper, we have modified the Rolie-Poly model \cite{Likhtman:2003} to incorporate Ianniruberto's flow-induced disentanglement theory and capture inhomogeneous disentanglement at the continuum level of the orientation tensor. Although the Rolie-Poly model handles stretch in a slightly different way to the molecular GLaMM model, the advantage of the Rolie-Poly model is the simple one-mode constitutive equation that can be applied to arbitrary inhomogeneous flows. We have shown that disentanglement during FFF is sensitive to the chosen CCR parameter $\beta$, particularly during the deposition stage due to the large stretch induced by the corner-flow geometry. 

Flow-induced disentanglement is observed in numerous simulations, including Brownian simulations \cite{Yaoita:2008}, molecular dynamics simulations \cite{Baig:2010} and dissipative particle dynamics simulations \cite{Khomami:2015}. 
To date $\beta$ serves as a fitting parameter between simulations and the tube theory, and different values are required depending on the flow type and entanglement number \cite{Ianniruberto:2015}. Thus, testing FFF-induced disentanglement using a range of constitutive models is a must. 

Various other CCR theories have arisen based on molecular simulations. For example, Wang \& Larson \cite{Wang:2008} incorporate kink-jump motions of the tube segments to capture the constraint release effect and find a broad distribution of constraint lifetimes. In contrast to tube models, slip-link models construct an effective field to represent entanglements so that the chain satisfies random-walk statistics at all length scales and constraint release is determined by a slip-link friction \cite{Schieber:2014, Likhtman:2005}. However, with little experimental evidence, flow-induced disentanglement continues to be a debated topic.

Lastly, the model presented here is restricted to predicting the behaviour of linear amorphous polymer melts. FFF systems can handle a wide range of rheologically complex materials, including semi-crystalline melts \cite{Drummer:2012} and filled melts, containing nano-scale particles, such as ABS \cite{Ziemian:2012}. Since semi-crystalline polymers tend to flow more readily compared to amorphous melts above the glass transition temperature, FFF systems incorporate fans that rapidly cool the extruded material \cite{Goyanes:2016}. Thus, it is expected that semi-crystalline printed parts will exhibit a greater degree of anisotropy than parts made from amorphous materials. 


Although the most-commonly-printed polymer is ABS, an amorphous melt containing rubber nano-particles that provide toughness even at low temperatures, this material has been rarely characterised rheologically \cite{Aoki:2001}. The addition of fibres to an amorphous melt can also enhance both thermal and mechanical properties. Yet how these fillers behave during the printing process and how they modify viscoelasticity remain open questions.

\begin{acknowledgments}
We thank Jonathan Seppala and Kalman Migler for advice and an enjoyable collaboration, as well as the National Institute for Standards and Technology (NIST), Georgetown University, and the Ives Foundation for funding. 
\end{acknowledgments}

\appendix
\section{The Constitutive Curve}
\label{sec:appendix1}

\begin{figure}[t!]
\centerline{\includegraphics[width=8.5cm]{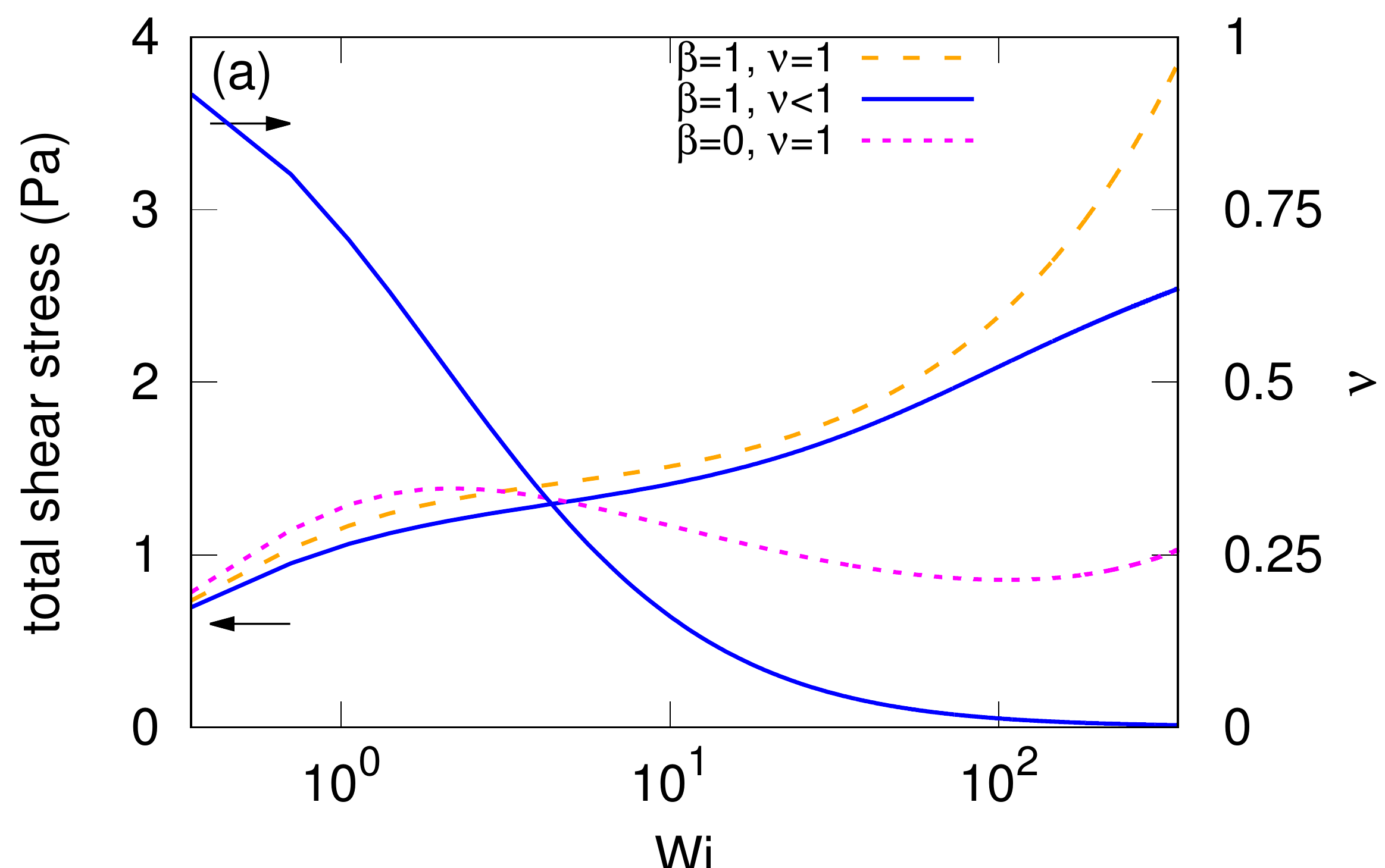}}
\centerline{\includegraphics[width=8.5cm]{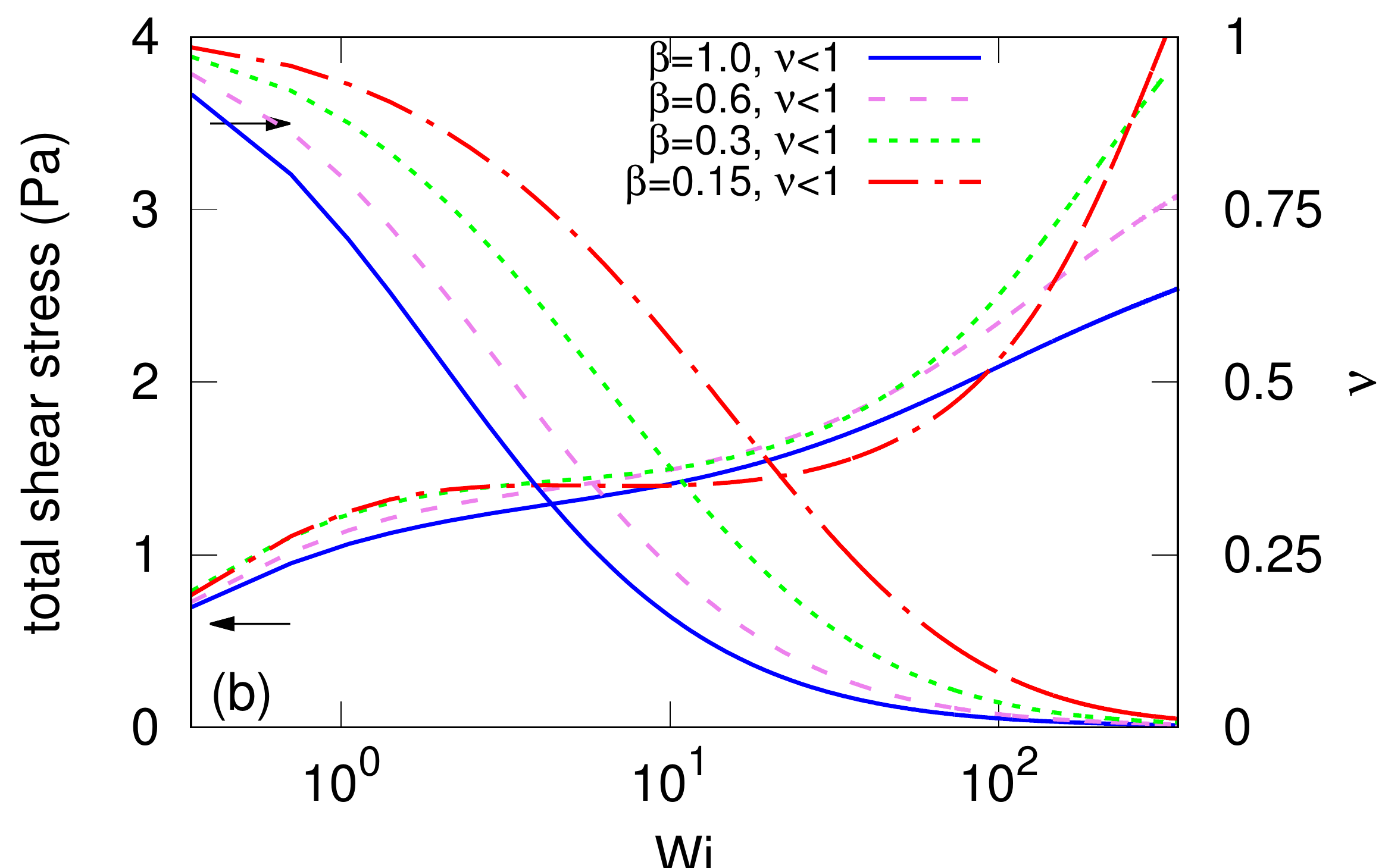}}
\caption{Log-linear plot of constitutive curve given by Eq. \ref{eq:stress} and the corresponding disentanglement fraction $\nu$ for $Z_{eq}=37$ and increasing Weissenberg number $Wi$; (a) effect of allowing disentanglement in Eq. \ref{eq:stress} via Eq. \ref{eq:nu} and (b) effect of changing CCR parameter $\beta$. }
\label{fig:Ccurve}
\end{figure}

The steady-state constitutive curve defined by Eqs. \ref{eq:stress}, \ref{eq:Rolie-Poly} and \ref{eq:nu} is plotted in Fig. \ref{fig:Ccurve}a for a $Z_{eq}=37$ melt that remains fully entangled $(\nu=1)$ and that is allowed to disentangle $(\nu < 1)$, where the value of $\nu$ varies with shear rate. The CCR parameter is set to $\beta = 0$ and 1. The curve indicates the shear-thinning nature of the Rolie-Poly model. Feed stocks for FFF processes are typically shear thinning and are often assumed to follow a power-law viscosity model \cite{Bellini:2004, Mostafa:2009, Ramanath:2008, Yardimici:1997}. These treatments are not molecularly aware and cannot capture normal stress effects of complex flow fields. The Rolie-Poly model, on the other hand, includes key aspects of the molecular melt structure. 

For $\beta=0$ there is no CCR so that $\nu$ can only equal unity in a steady flow and the constitutive curve is non-monotonic. For $\beta =1$, disentanglement can occur and the entanglement fraction $\nu$ becomes less than unity for sufficiently large shear rates. This disentanglement mechanism acts to suppress excess shear-thinning behaviour in a similar way to increasing the CCR parameter $\beta$ \cite{Ianniruberto:1996}, as is demonstrated in Fig. \ref{fig:Ccurve}b. Smaller $\beta$ flattens the constitutive curve, enhancing shear-thinning behaviour by reducing the rate of convective constraint release. For $\beta=0.15$ the constitutive curve becomes non-monotonic for $Z_{eq}=37$, in which case shear-banding instabilities would be expected in the nozzle \cite{Olmsted:2008}. 

\section{Full Deposition Calculation}
\label{sec:appendix3}

\subsection{Parametrisation}

To parametrise the shape of the curved filament, we deform a cylinder into an elliptic-cylinder by rotating successive planes, as shown in Fig. \ref{fig:mesh}a. We divide the space into a three-dimensional mesh, where each plane is specified in terms of the spherical coordinate system $(r,\theta,\phi)$, where $r$ defines the radial position from the centre of the plane and $\phi$ is the azimuthal angle around the plane. The angle between the nozzle and the fully-deposited layer is denoted by $\theta$ and is in the range $ \theta \in [0,\pi/2]$.  The initial plane (located at the nozzle exit) is a circle of radius $R$ centred at $(0,0,-H)$ with $\theta=0$ (Fig. \ref{fig:mesh}b). The final plane at $\theta = \pi/2$ is an ellipse centred at $(0,R,-H/2)$ with major radius $R$ and minor radius $H/2$ (Fig. \ref{fig:mesh}d).

The mesh points $\boldsymbol{\mathcal{R}}=(x,y,z)$ are expressed as Cartesian functions of the spherical coordinate system $(r,\theta,\phi)$. That is, in the frame moving with the nozzle,
\begin{equation}
 \boldsymbol{\mathcal{R}}(r,\theta,\phi) = x(r,\theta,\phi) \hat{\bf e}_x +  y (r,\theta,\phi) \hat{\bf e}_y + z (r,\theta,\phi) \hat{\bf e}_z.
\end{equation}
Coordinates for each plane are calculated by applying a deformation to the initial plane, parametrised by initial polar coordinates $(r^0, \phi^0)$, rotated by angle $\theta$ about the stagnation point at the nozzle exit. That is,
\begin{equation}
 \boldsymbol{\mathcal{R}}(r,\theta, \phi) =  \boldsymbol{\mathcal{T}}(\theta) \cdot \boldsymbol{\mathcal{R}}(r^0,0, \phi^0),
 \label{eq:mapping}
\end{equation}
where
\begin{equation}
\boldsymbol{\mathcal{T}}(\theta) \equiv {\bf \Lambda}(\theta) \cdot {\bf M}(\hat{x}, \theta).
\end{equation}
That is, a rotation about the $\hat{x}$-axis for angle $\theta$ defined by
\begin{equation}
 {\bf M}(\hat{x},\theta)  = \left( \begin{array}{ccc}
 1 & 0 &  0  \\
0 & \cos \theta & -\sin\theta \\
0 & \sin \theta & \cos \theta \\
\end{array} \right),
\end{equation}
and a deformation factor given by
\begin{equation}
 {\bf \Lambda} (\theta)= \left( \begin{array}{ccc}
 \lambda_x(\theta) & 0 &  0  \\
0 & \lambda_y(\theta) & 0 \\
0 & 0 & \lambda_z(\theta), \\
\end{array} \right).
\end{equation}
This is a general mapping for any function ${\bf \Lambda}$. In our case, the deformation is defined by
\begin{subequations}
 \begin{align}
\lambda_x = \lambda_y &= 1, \\
\lambda_z &= \frac{H}{2R},
\end{align}
\label{eq:smoothfactor}
\end{subequations}
so that the outer corner of the deposition shape traces an ellipse. Eq. \ref{eq:smoothfactor} defines the typical geometry transformation imposed by the FFF process; since the nozzle head is placed at height $H$ above the build plate (or previously-printed layer), which is less than the nozzle diameter $2R$, printed layers are elliptically shaped and mass is conserved by balancing $U_N$ and $U_L$ (Eq. \ref{eq:massconservation}).

\subsection{Initial condition}

The initial velocity is assumed to be uniform at the nozzle exit and is given by ${\bf u} = (0,0, U_N)$. The initial polymer configuration ${\bf A}$ induced by flow through the nozzle is calculated in Section \ref{sec:nozzle} and in cylindrical polar coordinates $(r^0, \phi^0, s)$. This is then converted to the the Cartesian frame $(x,y,z)$ to calculate the deposition flow. In the following $i,j,k ,\dots$ label Cartesian and $\alpha,\beta,\gamma, \dots$ label spherical polar coordinates. 

The polymer tensor $A_{\alpha \beta}$ is converted to Cartesian coordinates $(x,y,z)$ via the rotation
\begin{equation}
 A_{ij} = \Omega_{i\alpha} A_{\alpha \beta} \Omega_{\beta j},
\end{equation}
where the rotation matrix is given by
\begin{equation}
\Omega_{i \alpha} =  \left( \begin{array}{ccc}
\cos \phi^0 & - \sin \phi^0 & 0 \\
\sin \phi^0 & \cos \phi^0 & 0 \\
0 & 0 & 1 \end{array} \right)_{i\alpha},
\end{equation}
so that the initial polymer configuration for the deposition calculation is given by
\begin{subequations}
 \begin{align}
 A_{xx} &= A_{rr}, \\
 A_{yy} &= A_{rr}, \\
 A_{zz} &= A_{ss}, \\
 A_{xy} &= 0, \\
 A_{xz} &= \cos \phi^0 A_{rs}, \\
 A_{yz} &= \sin \phi^0 A_{rs},
\end{align}
\end{subequations}
since there is zero second normal stress ($A_{rr} = A_{\theta \theta}$) in the Rolie-Poly model under axisymmetric flow. 

\subsection{Mesh Spacing}

The initial geometry (polar angle $\theta=0$) is given by a circular plane defined in Cartesian coordinates by
\begin{equation}
 \boldsymbol{\mathcal{R}}(r^0,0,\phi^0) = r^0 \cos \phi^0 \ \hat{\bf e}_x + r^0 \sin \phi^0 \ \hat{\bf e}_y,
\end{equation}
for initial polar coordinates $(r^0,\phi^0)$. This initial plane is divided up into a numerical mesh (Fig. \ref{fig:mesh}b) with mesh spacing given by
\begin{equation}
\delta {\bf s} = \delta r^0 \ {\bf \hat{r}}^0 +  r^0 \delta \phi^0 \ \boldsymbol{\hat{\phi}}^0,
\label{eq:meshspacing_uniform}
\end{equation}
where 
\begin{equation}
  r^0 = |\boldsymbol{\mathcal{R}} - \boldsymbol{\mathcal{R}}^o|,
 \end{equation}
for plane centre $\boldsymbol{\mathcal{R}}^o ={\bf 0}$. For this initial circular geometry, the radial and azimuthal spacing, $\delta r^0$ and $\delta \phi^0$, are uniform and given by
\begin{equation}
 \delta r^0 = \frac{2R}{M_{max}} \quad \text{ and } \quad \delta \phi^0 = \frac{2\pi}{P_{max}},
\end{equation}
where $M_{max}$ and $P_{max}$ are the number of radial and azimuthal mesh points, respectively.

Due to the nature of the mapping, it is natural to continue adopting the parametrisation $(r,\theta,\phi)$ for subsequent planes. However, the initial polar coordinates $(r^0, \phi^0)$ defined for the circular plane, are not equivalent for subsequent planes.
In general, the coordinates
\begin{equation}
 r \equiv r(r^0, \theta, \phi^0) \quad \text{ and } \quad \phi \equiv \phi(r^0, \theta, \phi^0),
\end{equation}
depend on the shape of the plane at $\theta$, which is determined by deformation that Eq. \ref{eq:mapping} imposes onto a circular plane. Note that $r$ and $\phi$ are not explicitly required to calculate $\boldsymbol{\mathcal{R}}$, but act as counters to locate adjacent mesh points within a plane.

In the case discussed in this paper the deformation imposes elliptical geometry. Thus, for $\theta > 0$,
\begin{equation}
 r \ne r^0 \quad \text{ and } \quad  \phi \ne \phi^0,
\end{equation}
and the mesh spacing is not uniform across and around each plane:
\begin{equation}
 \delta r \ne \delta r^0 \quad \text{ and } \quad  \delta \phi \ne \delta \phi^0,
\end{equation}
Since the mesh spacing must now reflect the spacing between mesh points that are mapped using Eq. \ref{eq:mapping}, we must make the following distinctions. 

\begin{figure}[b!]
\centerline{\includegraphics[width=8cm]{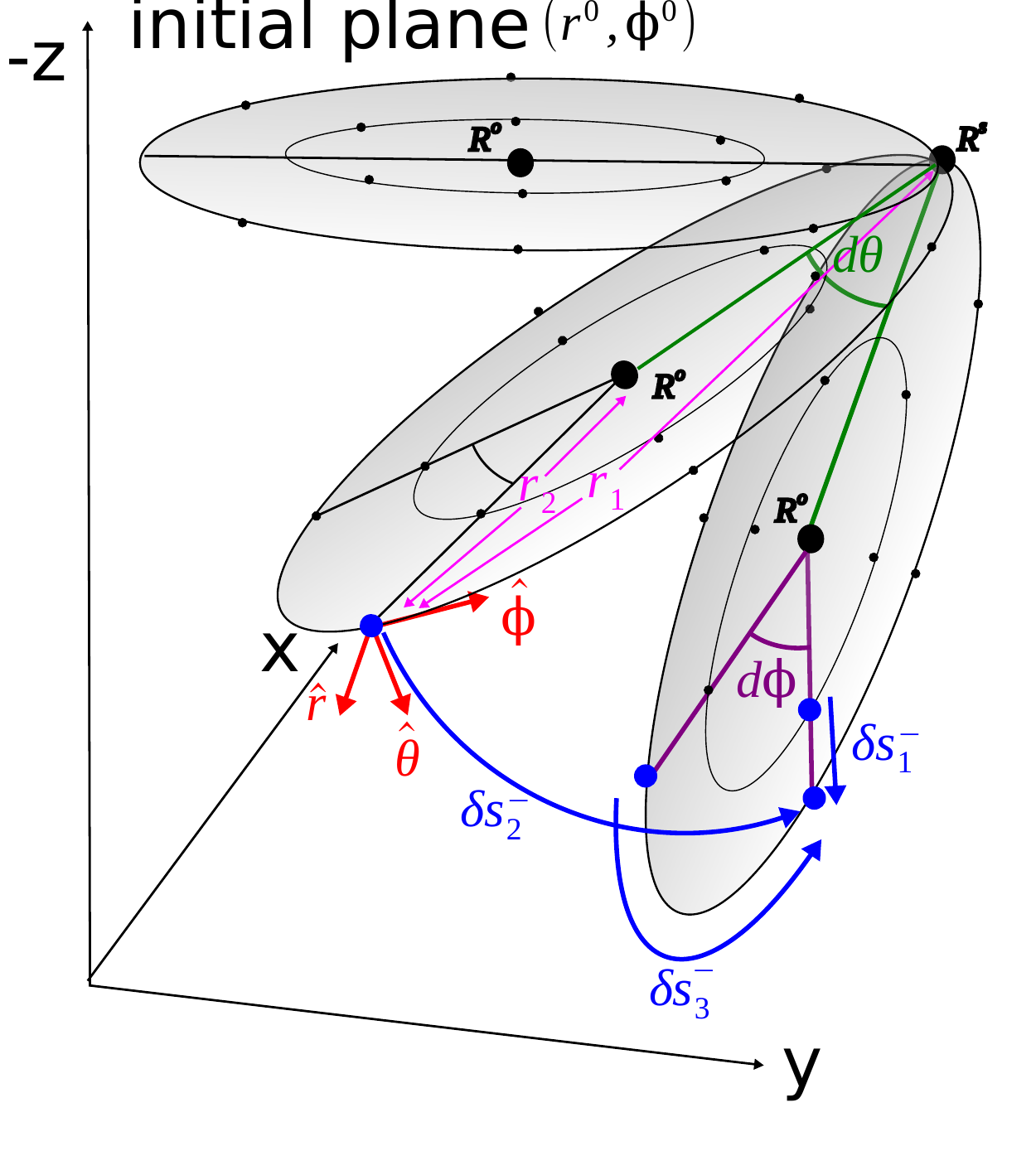}}
\caption{Schematic of numerical method showing the initial circular plane and two successive planes that are elliptic in shape. Cartesian mesh points are parametrised by spherical polar coordinates $(r,\theta,\phi)$. Mesh spacing $\delta s_\alpha^-$ is given by Eq. \ref{eq:meshspacing}. Note that the radial spacing $\delta r$ and the azimuthal angle spacing $\delta \phi$ are not uniform around and across each plane for $\theta > 0$ due to the elliptical geometry. See text for details. }
\label{fig:drawing}
\end{figure}

First, the mesh spacing is given by
\begin{equation}
\delta {\bf s}^\pm = \delta r \ {\bf \hat{r}} +  r_1 \delta \theta \ \boldsymbol{\hat{\theta}} + r_2 \delta \phi \ \boldsymbol{\hat{\phi}} ,
\label{eq:meshspacing}
\end{equation}
where $\pm$ signifies the forward and backward directions for each coordinate. Second, the arc lengths in Eq. \ref{eq:meshspacing} are calculated from the average of two successive arcs by defining
\begin{subequations}
\begin{align}
r_1^\pm  &\equiv \frac{1}{2} \Bigg( \vert \boldsymbol{\mathcal{R}}-\boldsymbol{\mathcal{R}}^s \vert + \vert \boldsymbol{\mathcal{R}}(r,\theta \pm \delta \theta,\phi)- \boldsymbol{\mathcal{R}}^s \vert \Bigg), \\
r_2^\pm &\equiv \frac{1}{2} \Bigg( \vert \boldsymbol{\mathcal{R}}-\boldsymbol{\mathcal{R}}^o \vert + \vert \boldsymbol{\mathcal{R}} (r, \theta, \phi \pm \delta \phi)-\boldsymbol{\mathcal{R}}^o \vert \Bigg),
\label{eq:radii}
\end{align}
\end{subequations}
where
\begin{equation}
 \boldsymbol{\mathcal{R}}^s \equiv (0,R ,0) \quad  \text{ and } \quad \boldsymbol{\mathcal{R}}^o \equiv \boldsymbol{\mathcal{T}}(\theta) \cdot \boldsymbol{\mathcal{R}}(0,0,0),
\end{equation}
are the stagnation point and centre of each plane. Finally, the radial spacing $\delta r$ is given by
\begin{equation}
 \delta r^\pm = \vert \boldsymbol{\mathcal{R}} - \boldsymbol{\mathcal{R}}(r \pm \delta r,\theta,\phi) \vert,
\end{equation}
the azimuthal angle $\delta \phi$ is calculated via the law of cosines
\begin{widetext}
\begin{equation}
 \cos \delta \phi^\pm = \frac{ \vert \boldsymbol{\mathcal{R}}-\boldsymbol{\mathcal{R}}^o \vert^2 + \vert \boldsymbol{\mathcal{R}} (r, \theta, \phi \pm \delta \phi)-\boldsymbol{\mathcal{R}}^o \vert^2 - \vert \boldsymbol{\mathcal{R}} - \boldsymbol{\mathcal{R}} (r, \theta, \phi \pm \delta \phi) \vert^2 } {2 \vert \boldsymbol{\mathcal{R}}-\boldsymbol{\mathcal{R}}^o \vert \vert \boldsymbol{\mathcal{R}} (r, \theta, \phi \pm \delta \phi)-\boldsymbol{\mathcal{R}}^o \vert }.
\end{equation}
\end{widetext}
and the polar angle is chosen to vary uniformly according to
\begin{equation}
 \delta \theta = \frac{\pi}{2 N_{max} },
\end{equation}
where $N_{max}$ is the total number of planes. The mesh spacing defined in Eq. \ref{eq:meshspacing} is shown in Fig. \ref{fig:drawing}.

\subsection{Flow Field}

The velocity profile is also written as Cartesian functions of the spherical coordinate system $(r, \theta, \phi)$, such that
\begin{equation}
\begin{split}
 {\bf u}(r,\theta,\phi) 
 &= U(r,\theta, \phi) \ \hat{\bf s}(\theta), \\
 &= v(r,\theta, \phi) \hat{\bf e}_y + w(r,\theta, \phi) \hat{\bf e}_z, 
 \end{split}
\end{equation}
where
\begin{equation}
 U = \sqrt{v^2 + w^2} \quad \text{and} \quad \hat{\bf s} = \sin\theta \hat{\bf e}_y + \cos \theta \hat{\bf e}_z,
\end{equation}
are the magnitude of the velocity and the local unit normal vector (normal to the plane), respectively. On exiting the nozzle, $\theta=0$ and ${\bf u} = (0,0,U_N)$, whereas at the end of deposition $\theta=\pi/2$ and the velocity profile of the layer is ${\bf u}= (0,U_L,0)$. 

Instead of solving the full Navier-Stokes equations, we assume $\tau_{dep} \ll \tau_d, \tau_R$. Thus, assuming no secondary flows, the velocity profile is calculated from the local flux-conservation condition 
\begin{equation}
 U(r, \theta, \phi) \ da(r,\theta, \phi) = U_N \ da(r^0,0, \phi^0),
\end{equation}
where $d a$ denotes the area of a single mesh element imposed by the prescribed shape (\ref{eq:mapping}), as shown in Fig. \ref{fig:mesh}b,c. 
Thus, the horizontal and vertical velocity components are given by
\begin{subequations}
\begin{align}
 v(r,\theta, \phi) &= U_N \frac{da(r^0,0, \phi^0)}{da(r,\theta, \phi)} \sin \theta, \\ 
 w(r,\theta, \phi) &= U_N \frac{da(r^0,0, \phi^0)}{da(r,\theta, \phi)} \cos \theta,
  \end{align}
  \label{eq:depvel}
\end{subequations}
respectively.

\subsection{Polymer Deformation}

Since the flow starts in steady state and polymer relaxation is ignored, the deposition flow remains in steady state. Thus, to advect the polymer with velocity gradients during deposition, we solve
\begin{equation}
 ({\bf u} \cdot \nabla) {\bf A} = {\bf K} \cdot {\bf A} + {\bf A } \cdot {\bf K}^T,
 \label{eq:advect}
\end{equation}
in the Cartesian frame. In Einstein notation, Eq. \ref{eq:advect} is written as
\begin{equation}
\begin{split}
\left( u_i  \partial_i \right) A_{jk} &=  K_{jl} A_{lk} + A_{jl} K_{kl}, \\ 
&= \partial_l u_j A_{lk} + A_{jl} \partial_l u_k.
\label{eq:adCart}
\end{split}
\end{equation}
where derivatives in the Cartesian frame are denoted
\begin{equation}
 \partial_i = \left( \frac{\partial}{\partial x}, \frac{\partial}{\partial y}, \frac{\partial}{\partial z}\right)_i.
\end{equation}
Since the Cartesian mesh points are defined as functions of the spherical polar coordinate system $(r,\theta,\phi)$, derivatives in the ${\bf \hat{r}}, \boldsymbol{\hat{\theta}}$ and $\boldsymbol{\hat{\phi}}$ directions, that is
\begin{equation}
 \partial_\alpha = \left( \frac{\partial}{\partial r}, \frac{1}{r_1} \frac{\partial}{\partial \theta}, \frac{1}{r_2} \frac{\partial}{\partial \phi}\right)_\alpha,
\end{equation}
are easily computed from the mesh, as shown in Fig. \ref{fig:drawing}.

Due to the non-uniform nature of the mesh, velocity gradients in the ${\bf \hat{r}}, \boldsymbol{\hat{\phi}}$ and $\boldsymbol{\hat{\theta}}$ directions are given by the average of a forward and backward first-order finite-difference approximation. For example, velocity gradients in the ${\bf \hat{r}}$-direction are given by
\begin{equation}
\begin{split}
\partial_r u_i  &= \frac{1}{2} \left( \frac{u_i(r+\delta r^+,\theta,\phi)-u_i}{\delta r^+}\right) \\ 
               &+ \frac{1}{2} \left(\frac{u_i-u_i(r-\delta r^-,\theta,\phi)}{\delta r^-} \right),
\end{split}
\label{eq:FD}
\end{equation}
and similarly for the $\boldsymbol{\hat{\theta}}$ and $\boldsymbol{\hat{\phi}}$-directions. One-sided finite-difference approximations are used at the boundaries.

Derivatives are converted to the Cartesian frame via
\begin{equation}
 \partial_\alpha = \mathcal{M}_{\alpha i} \partial_i.
\end{equation}
where 
\begin{equation}
 \mathcal{M}_{\alpha i} = \left( \begin{array}{ccc}
 \cos\phi & \cos\theta\sin\phi &  -\sin\theta\sin\phi  \\
 0 & \sin \theta & \cos \theta \\
-\sin \phi & \cos \theta \cos \phi & -\sin\theta\cos\phi \\
\end{array} \right)_{\alpha i}.
\end{equation}
Thus, Eq. \ref{eq:adCart} becomes
\begin{equation}
u_i \mathcal{M}_{\alpha i}^{-1} \partial_\alpha A_{jk} = \mathcal{M}_{\alpha l}^{-1} \partial_\alpha u_j A_{lk} + A_{jl} \mathcal{M}_{ \alpha l}^{-1} \partial_\alpha u_k ,
\end{equation}
and tensor $A_{jk}$ is advected in the $\boldsymbol{\hat{\theta}}$-direction through angle $\theta$ via
\begin{equation}
 \Gamma_2 \partial_2 A_{jk} = \left( \mathcal{M}_{\alpha j}^{-1} \partial_\alpha u_l A_{lk} + A_{jl} \mathcal{M}_{\alpha k}^{-1} \partial_\alpha u_l \right) - \sum_{\alpha=1,3} \Gamma_\alpha \partial_\alpha A_{jk},
\end{equation}
where
\begin{equation}
\Gamma_\alpha = u_i \mathcal{M}_{\alpha i}^{-1} .
\end{equation}
We make the forward finite-difference approximation
\begin{equation}
\Gamma_2 \frac{1}{r_1^+} \frac{\partial A_{jk}}{\partial \theta}  = \frac{A_{jk}(r,\theta+\delta\theta,\phi) - A_{jk}}{\Delta t},
\end{equation}
where 
\begin{equation}
 \Delta t = \frac{r_1^+\delta \theta}{\Gamma_2},
 \label{eq:advectiontime}
\end{equation}
is the advection time scale that captures a greater displacement $ds = r_1^+ \delta \theta$ at the outside edge of the deposition (Fig. \ref{fig:visualdep}a).

Then, the semi-implicit finite-difference scheme is defined by the generalised matrix system
\vspace{1cm}
\begin{widetext}
\begin{equation}
\Bigg[ A_{jk} - \Delta t \left(
  \mathcal{M}_{\alpha l}^{-1} \partial_\alpha u_j A_{lk} + A_{jl} \mathcal{M}_{\alpha l}^{-1} \partial_\alpha u_k \right)
 \Bigg]_{(r,\theta+\delta \theta,\phi)}  = \quad
\Bigg[ A_{jk} + \Delta t \left( \mathcal{M}_{\alpha l}^{-1} \partial_\alpha u_j A_{lk} + A_{jl} \mathcal{M}_{ \alpha l}^{-1} \partial_\alpha u_k \right) - \Delta t \sum_{\alpha=1,3} \Gamma_\alpha \partial_\alpha A_{jk} .\Bigg]_{(r,\theta,\phi)},
\label{eq:implicitscheme}
 \end{equation}
\end{widetext}
to be solved for $A_{jk}(r,\phi)$ on the plane at angle $\theta + \delta \theta$ using information from the previous plane at angle $\theta$. Note that $M_{\alpha l}$ and $u_j$ are known for all planes a priori based on the transformation given by Eq. \ref{eq:mapping} and the flux-conservation condition Eq. \ref{eq:depvel}, respectively. The final results are given in the flow coordinate system via
\begin{equation}
 A_{\alpha \beta} = \mathcal{M}_{\alpha i} A_{ij} \mathcal{M}_{j\beta}.
\end{equation}

\section{Approximations and Assumptions in the Model}
\label{sec:appendix2}

\begin{table*}[pt]
\caption{Model parameters for typical amorphous printing material, polycarbonate.}
\centerline{\begin{tabular}{l|c|c|c}
\hline
\hline
 {\bf Polycarbonate Properties} & Notation & Value & Units \\
\hline
\hline
{Reference Temperature} & $T_0$ & 260& $^\text{o}$C \\
{Thermal Diffusivity \cite{Zhang:2002}} (at 25$^\text{o}$C) & $\alpha$ & 0.144 & mm$^2$/s \\
\hline
{Molecular Weight} & $M_w$ & 60 & kDa \\
{Entanglement Molecular Weight \cite{Mark:1996}} & $M_e$ & 1.6 & kDa \\
{Plateau Modulus \cite{Mark:1996} } & $G_e$ & $2.6 \times 10^6$ & Pa \\
{Entanglement Time (at $T_0$)} \cite{DoiEdwards:1986} &$ \tau_e^0$ & $3.29 \times 10^{-7}$  & s \\
\hline
{WLF parameter } & $C_1$ & 3  & - \\
{WLF parameter } & $C_2$ & 160 & - \\
\hline
{Equilibrium Entanglement Number} & $Z_{eq}$ & 37 & - \\
{Equilibrium Reptation Time} (at $T_N$) Eq. \ref{eq:taudeq} & $ \tau_d^{eq}$ & 0.03 & s \\
{Equilibrium Rouse Time} (at $T_N$) Eq. \ref{eq:tauReq} & $\tau_R^{eq}$ & $5.7 \times 10^{-4}$ & s \\
\hline
\end{tabular}}
\label{tab:polycarbonate}

\caption{Model parameters for two typical print speeds corresponding to a `fast' and `slow' case.}
\centerline{\begin{tabular}{l|c|c|c|c}
\hline
\hline
 {\bf Printing Parameters} & Notation & Fast Case & Slow Case & Units \\
\hline
\hline
{Mass flow rate} &  $Q$                     & $9.29 \times 10^{-6}$ & $1.26 \times 10^{-6}$   & kg/s\\
\hline
{Mean Initial Speed} (heated nozzle section)  &  $U_0$          & 3 & 0.5 & mm/s\\
{Mean Extrusion Speed} (final nozzle section) & $U_N$                          & 75 & 10      &  mm/s     \\
{Mean Print Speed} (across deposited layer) & $U_L$                              & 100 & 13   &  mm/s  \\
\hline
{Thermal Diffusion Time} (heated nozzle section)     Eq. \ref{eq:thermaldiffusivity}    & $\tau_\alpha$ & 7 & 7 & s \\
{Residence Time} (heated nozzle section) Eq. \ref{eq:restime0} & $\tau_{res}^0$        & 2 & 12 & s \\
\hline
{Nozzle Temperature} & $T_N$ & 250 & 250& $^\text{o}$C \\
{Deposition Time} Eq. \ref{eq:deptime}                & $\tau_{dep}$           & 0.005 & 0.03 & s \\
Thermal Skin Layer in Deposit Eq. \ref{eq:skinlayer} & $L_{skin}$ & 0.06 & 0.16 & mm \\
\hline
{Residence Time} (final nozzle section) Eq. \ref{eq:restime} & $\tau_{res}$           & 0.011  &   0.08 & s     \\
\hline
Die Swell Time  & $\tau_{sw}$ & 0.0352 & 0.0352&  s \\
Terminal Swell Distance Eq. \ref{eq:swelldistance} & $z_M$ & 2.64 & 0.352&  mm \\
\hline
\end{tabular}}
\label{tab:speeds}

\caption{Model parameters for typical nozzle geometry, as shown in Fig. \ref{fig:process}.}
\centerline{\begin{tabular}{l|c|c|c}
\hline
\hline
 {\bf Nozzle Dimensions} & Notation & Value & Units \\
\hline
\hline
Temperature & $T_N$ & 250 & $^\text{o}$C \\
\hline
{Radius} (heated nozzle section) & $R_0$ & 1.0 & mm \\
{Length} (heated nozzle section) & $L_0$ & 6.0 & mm \\
\hline
{Radius} (final nozzle section) & $R$ & 0.2 & mm \\
{Length} (final nozzle section) & $L$ & 0.8 & mm \\
\hline
{Layer Thickness} & $H$ & 0.3 & mm \\
\hline
\end{tabular}}
\label{tab:geometry}
\end{table*}


\begin{figure}[t]
\begin{minipage}[t]{8cm}
\centerline{\includegraphics[width=6cm]{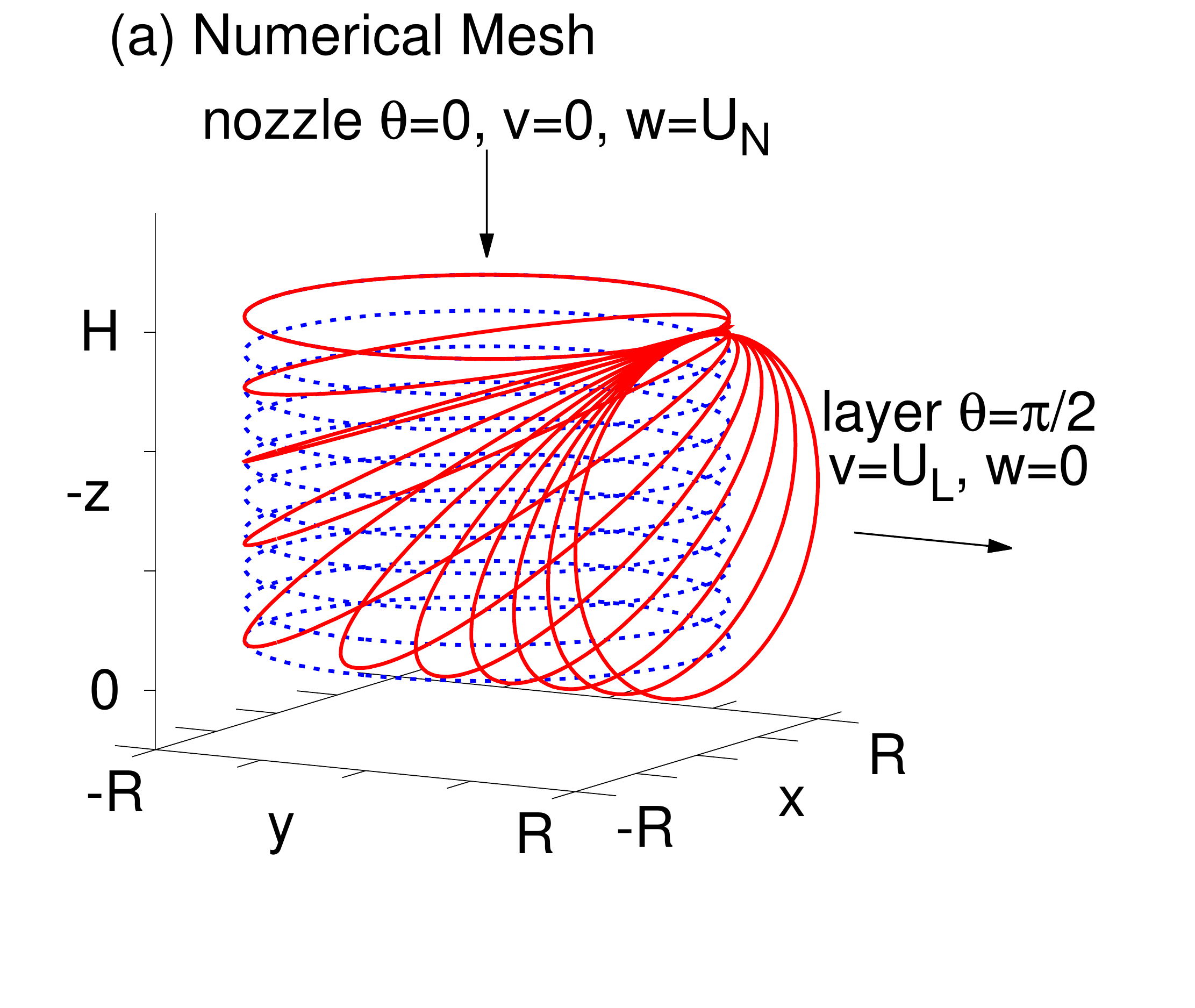}}
\end{minipage}
\begin{minipage}[t]{4cm}
 \centerline{\includegraphics[width=4cm]{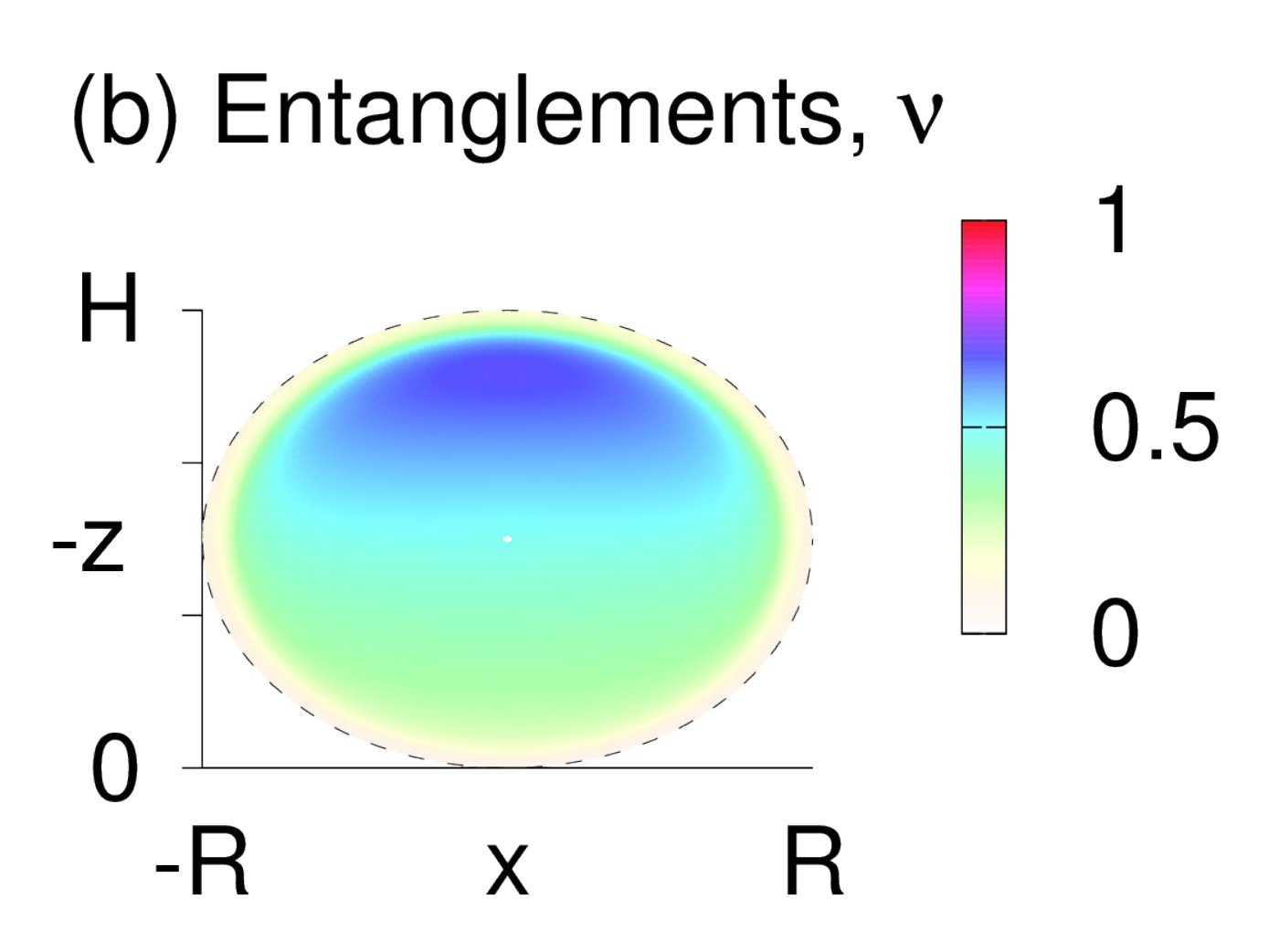}}
\end{minipage}
\begin{minipage}[t]{4cm}
 \centerline{\includegraphics[width=4cm]{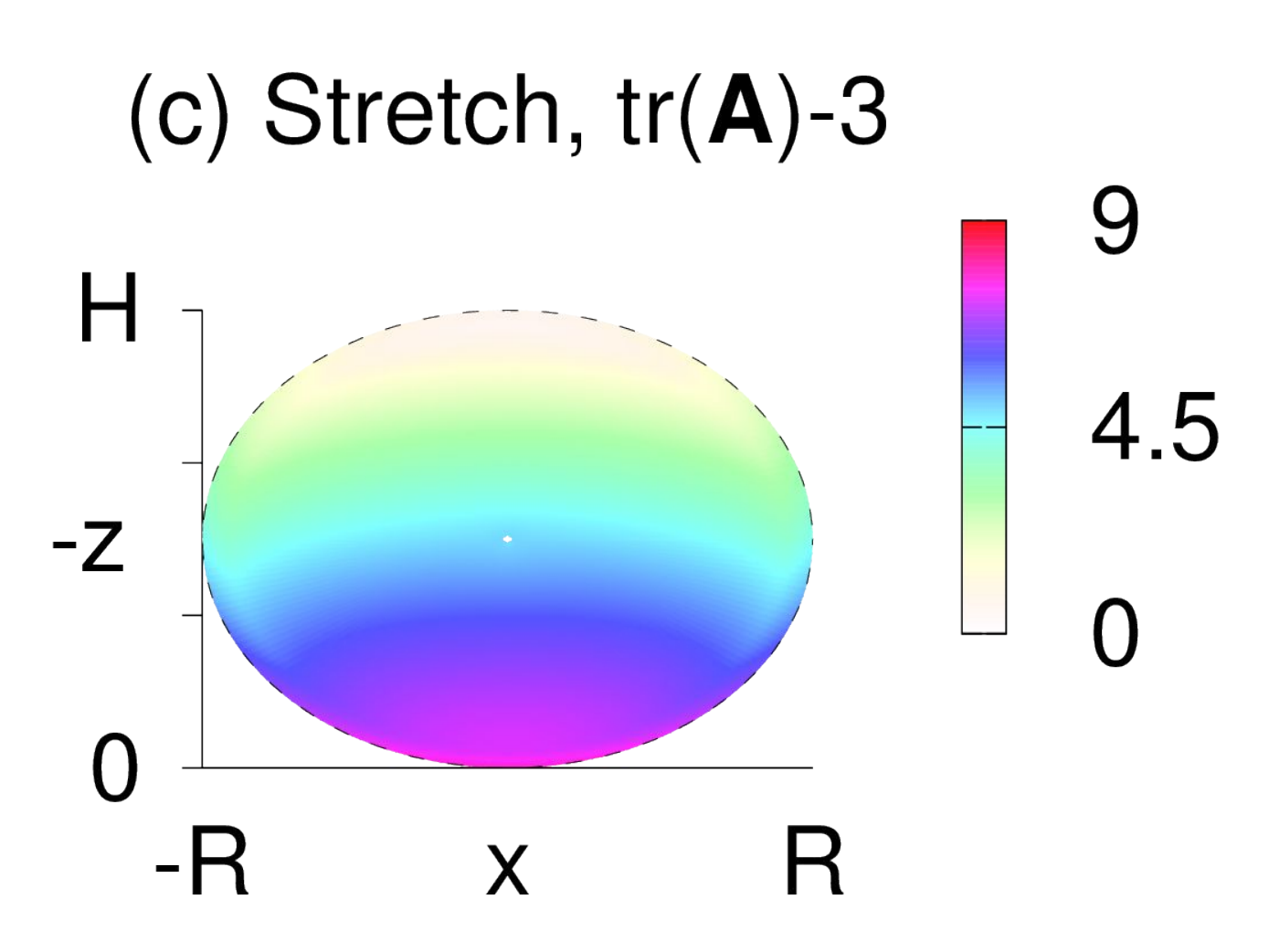}}
\end{minipage}
\caption{(a) Numerical mesh with a square corner at $(y,z)=(-R,0)$ and resulting polymer deformation across the printed layer: (b) entanglement $\nu(r,\phi)$ and (c)  Tube stretch profile $\text{tr}{\bf A}(r,\phi)-3$ for $Z_{eq}=37$, $\beta =0.3$ and $\overline{Wi}_N=2$. The blue dotted lines show the equivalent cylindrical mesh with the same volume.}
\label{fig:square}
\end{figure}

\begin{figure}[t]
\centerline{\includegraphics[width=8.5cm]{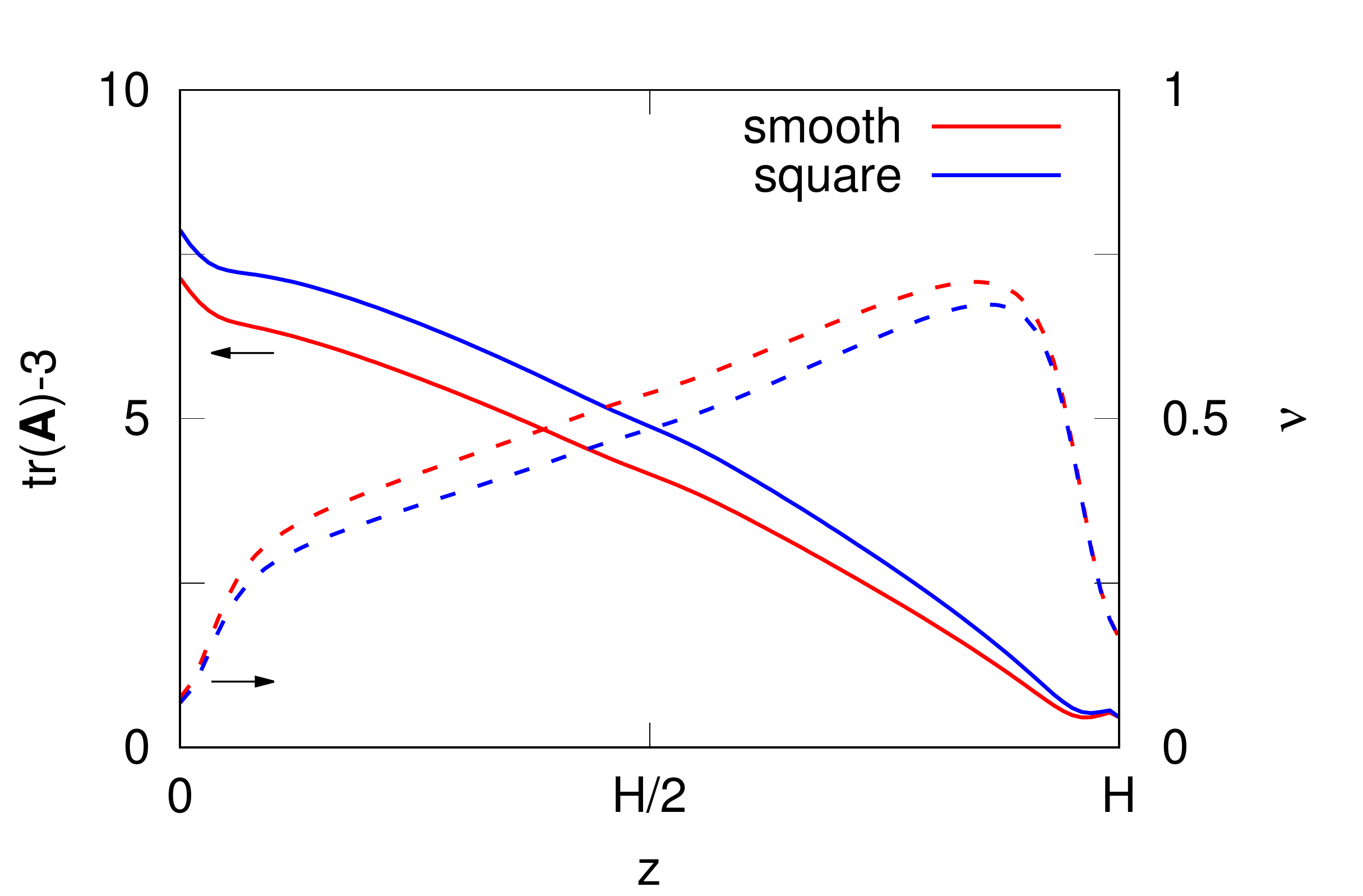}}
\caption{Quantitative comparison of the stretch $\text{tr}({\bf A})$ {\em (solid lines)} and disentanglement $\nu$ {\em (dashed lines)} along the $z$-axis ($(x,y)=(0,R)$) for a deposition shape with a smooth corner (Eq. \ref{eq:smoothfactor}, Fig. \ref{fig:mesh}a) and a square corner (Eq. \ref{eq:squarefactor}, Fig. \ref{fig:square}). Model parameters are $Z_{eq}=37$, $\beta =0.3$ and $\overline{Wi}_N=2$. }
\label{fig:shape_quant}
\end{figure}

Here we detail typical parameters and discuss further details of the validity of the FFF model. The model parameters for polycarbonate are given in Table \ref{tab:polycarbonate}, and typical print speeds and nozzle dimensions (corresponding to the simplified schematic in Fig. \ref{fig:process}) are given in Tables \ref{tab:speeds} and \ref{tab:geometry}, respectively. The assumptions made are as follows:

\begin{enumerate}
 \item We ignore viscous heating and assume that there is zero temperature gradient across the nozzle radius. A thick filament of solid polycarbonate is fed into the FFF nozzle at a mass flow rate $Q$. As the solid filament enters the nozzle, a heated element increases the temperature of the material to $T_N$ so that it becomes molten and flows at average speed $U_0$ such that
\begin{equation}
 Q = \rho \pi R_0^2 U_0,
\end{equation}
for density $\rho$ and nozzle radius $R_0$. The time scale for heat diffusion is given by 
\begin{equation}
\tau_\alpha =R_0^2/\alpha ,
\label{eq:thermaldiffusivity}
\end{equation}
for thermal diffusivity $\alpha$. If the residence time in the heated section of the nozzle satisfies 
\begin{equation}
\tau_{res}^0 = L_0/U_0 \gg \tau_\alpha \approx 7 \text{ s},
\label{eq:restime0}
\end{equation}
for length $L_0$, then the temperature is expected to be uniform across nozzle radius.  Finite-element analyses of thermal diffusion in the heated nozzle section (which ignore viscous heating effects) show that uniform temperature profiles $(\Delta T \le 1^\text{o}C)$ are rapidly achieved \cite{Bellini:2004,Mostafa:2009,Ramanath:2008}. 
For our model parameters $\tau_{res}^0 \sim \tau_{\alpha}$ for typical print speeds (Table \ref{tab:speeds}), so effects of a non-uniform temperature profile within the nozzle may need to be considered.

\item We assume that the temperature throughout the deposition is uniform. A simple calculation from thermal diffusivity suggests a cool boundary layer near the free surface of thickness
\begin{equation}
 L_{skin} = \sqrt{\tau_{dep} \alpha},
 \label{eq:skinlayer}
\end{equation}
which depends on the print speed through the deposition time; for a typically fast print speed $L_{skin}=0.06$ mm (Table \ref{tab:speeds}). We choose to neglect the complicated heat transfer process here.

\item We assume that the flow is steady state in the nozzle and during deposition. The molten material exits the final nozzle section at average speed $U_N$ (usually after passing through two contractions). If the residence time in the final nozzle section satisfies
\begin{equation}
\tau_{res} = L/U_N \gg \tau_d^{eq},
\label{eq:restime}
\end{equation}
for length $L$, then the flow can be assumed to be steady. For polycarbonate rheology, $\tau_{res} \sim \tau_d^{eq}$ for typical printing speeds (Table \ref{tab:speeds}), thus a more detailed calculation of the flow may be required to capture start-up effects in the nozzle. 

 
\item We assume that the time scale for die swell to fully develop is larger than the deposition time scale. Upon exiting the nozzle, since the melt is no longer constrained, the polymer conformations relax and elastically stored energy is released leading to die swell. The die-swell ratio $D_M/2R$, where $D_M$ denotes the maximum steady-state diameter of the melt after exiting the nozzle, is estimated from the first normal stress difference and the shear stress at the wall \cite{Tanner:1970}. Using the values calculated in section \ref{sec:nozzle}, the die-swell ratio is found to be \cite{Tanner:1970}
\begin{widetext}
\begin{equation}
 \frac{D_M}{2R} = \left( 1 + \frac{1}{2} \left( \frac{A_{ss}-A_{rr}}{2A_{rs}} \right)_W \right)^{1/6} + 0.13 \approx 1.2,
\end{equation}
\end{widetext}
for $\overline{Wi}_N=2$ and 13. Reported values for FFF-like processes range from $\sim$ 1.05 to 1.3 \cite{Turner:2014} and this swelling phenomenon is found the affect the alignment of extruded fibre suspensions \cite{Heller:2016}. Ceramic particles \cite{Bellini:thesis} and carbon fibres \cite{Shofner:2003} may be used to reduce the swelling effect.

\item We prescribe the shape of the deposition and neglect any spreading of the deposition on the build plate. Models that address the spreading of a printed layer \citep{Crockett:1996, Crockett:thesis} have yet to be applied to polymer melts. The deposition shape described in Section \ref{sec:deposition} assumes a smooth corner region (Fig. \ref{fig:mesh}a). However, the prescribed shape can effect the deformation imposed by the deposition process. As a comparison we have calculated the effect of having a sharp corner, whose deformation is defined by
\begin{subequations}
\label{eq:squarefactor}
\begin{align}
\lambda_x &= 1, \\
\lambda_y &=\lambda_z = \begin{cases} \displaystyle{ \frac{1}{\cos\theta}}, \text{ for } \theta < \theta^*, \\[10truept]
                                       \displaystyle{ \frac{\tan \theta^*}{\sin \theta}}, \text{ for } \theta > \theta^{*},
                                       \end{cases}
\end{align}
\end{subequations}
where $\theta^* = \tan^{-1}(R/H) $ denotes the angle at which the corner is reached (Figure \ref{fig:square}a). In this way the total deposited volume equals that of the cylinder that would be deposited during vertical extrusion with no die swell.  
Fig. \ref{fig:square}b,c shows that qualitatively the stretch and disentanglement profiles across the layer are similar to the smooth corner case (Fig. \ref{fig:deposition}b). A quantitative comparison is shown in Fig. \ref{fig:shape_quant}. A square outer-corner region induces more stretch and disentangles the melt further; at $z=0$ the stretch increases from $\text{tr}{\bf A}= 10.13$ to $10.87$ (approximately 93\% smaller with a smooth corner) and $\nu$ decreases from $\nu_L =0.0075$ to $0.0068$ (approximately 10\% larger with a smooth corner).

\end{enumerate}

\bibliography{references}

\end{document}